\documentclass[a4paper,12pt]{article}
\usepackage{t1enc}
\usepackage[english]{babel}
\usepackage[utf8]{inputenc}
\usepackage{graphicx}
\usepackage{amsmath}
\usepackage{amssymb}
\usepackage{hyperref}

\def\d{\mathrm{d}}

\def\lag{{\mathcal{L}}}
\def\kihagy#1{}

\newcommand{\be}{\begin{equation}}
\newcommand{\ee}{\end{equation}}
\newcommand{\bea}{\begin{eqnarray}}
\newcommand{\eea}{\end{eqnarray}}

\def\pa{\partial}

\title{Instabilities of Twisted Strings}

\medskip

\author{
P\'eter Forg\'acs$^{1,2}$, \'Arp\'ad Luk\'acs$^1$\\
\medskip\\
\small \sl $^1$MTA RMKI, H-1525 Budapest 114, P.O.Box 49, Hungary,\\
\small \sl $^2$LMPT, CNRS-UMR 6083, Universit\'e de Tours,\\
\small \sl Parc de Grandmont, 37200 Tours, France
}
\date{}

\begin{document}
\maketitle

\begin{abstract}
A linear stability analysis of {\sl twisted} flux-tubes (strings) in an SU(2) semilocal theory --
an Abelian-Higgs model with {\sl two} charged scalar fields with a global SU(2) symmetry --
is carried out. Here the twist refers to a relative phase between the two complex scalars
(with linear dependence on, say, the $z$ coordinate), and
importantly it leads to a global current flowing along the the string.
Such twisted strings bifurcate with the Abrikosov-Nielsen-Olesen (ANO) solution
embedded in the semilocal theory.
Our numerical investigations of the small fluctuation spectrum confirm 
previous results
that twisted strings 
exhibit instabilities whose amplitudes grow exponentially in time.
More precisely twisted strings with a single magnetic flux quantum
admit a continuous family of unstable eigenmodes with harmonic $z$ dependence,
indexed by a wavenumber $k\in[-k_{\rm m},k_{\rm m}]$.
Carrying out a perturbative
semi-analytic analysis of the bifurcation, it is found that the purely numerical results are very well
reproduced. 
This way one obtains not only a good qualitative
description of the twisted solutions themselves as well as of their instabilities,
but also a quantitative description of the numerical results.
Our semi-analytic results indicate that 
in close analogy to the known instability of the embedded ANO vortex
a twisted string is also likely to expand in size caused by the spreading 
out of its magnetic flux. 
\end{abstract}
{\small
\begin{flushleft}
\end{flushleft}
}

\section*{Introduction}

Topological defects, such as domain walls, strings and monopoles
arise in many field theory models with spontaneous symmetry breaking.  
These objects are expected to be created during phase transitions,
and are likely to play an important role in the early universe and more generally in cosmology
\cite{cs-review}.
Typically these defect solutions are stable because there is
an infinite energy barrier separating them
from the vacuum. An important class of defects is constituted by 
line-defects called cosmic strings. A cosmic string is typically a flux-tube,
enclosing a certain number of magnetic flux quanta.
In the plane orthogonal to its direction
a cosmic string corresponds to a vortex solution. 
Vice versa, any vortex in the plane can be extended to a straight string
in the orthogonal direction to the plane.
The 
well-known Abrikosov-Nielsen-Olesen (ANO) vortex of the Abelian Higgs model
serves as a prototype straight cosmic string.
ANO vortices are characterized by an integer winding number, $n$, determining their magnetic
flux, and by the mass ratio, $\beta=m_{\rm s}^2/m_{\rm v}^2$
($m_{\rm s}$ resp.\ $m_{\rm v}$
denoting the mass of the scalar resp.\ vector fields).

Rather interesting, cosmic string-type defect solutions have been found in 
semilocal theories where there is no topological stability \cite{vac-ach,hin,Vach91}.
An important class of semilocal theories is provided by Abelian Higgs models with a
suitably extended scalar sector,
the simplest one being two complex scalar fields admitting global SU(2) symmetry.
Semilocal models are quite interesting since
both global and local symmetries are simultaneously broken; for
a comprehensive review we refer to \cite{semilocal}.
Abelian Higgs models with an extended scalar sector contain ANO-type vortices, which correspond to
a simple embedding. 
Quite remarkably semilocal models exhibit stable vortices,
despite the first homotopy group of the vacuum manifold being trivial
\cite{vac-ach,hin}.
The linear stability of ANO vortices embedded into SU(N) symmetric semilocal models
has been examined in Ref.\ \cite{hin}, by computing their small fluctuation spectrum.
It has been found that embedded ANO vortices are stable only if
$\beta\leq1$. For $\beta > 1$ there is a family of energy lowering eigenmodes, 
parameterized by the $z$-direction wave number, 
$k \in [-k_{\rm m}, k_{\rm m}]$, with the lowest lying eigenmode
being $z$-independent. 
This homogeneous unstable mode can be seen to correspond to a ``magnetic spreading''
instability \cite{hin, Preskill}.
The long time dynamics of the instability (i.e.\ the numerical solution
of the full nonlinear field equations) has also been studied, in Ref.\ \cite{akpv}. 
There it is found that for $\beta >1$ embedded ANO vortices
undergo indeed a homogeneous expansion as expected from the analysis of Ref.\ \cite{hin}.

The case $\beta=1$ is quite special, for here,
instead of there being a unique (and stable) vortex solution satisfying
the first order Bogomol'nyi equations \cite{bog},
there is a continuous family of them having the same energy  \cite{hin}.
The Standard Model counterparts of the embedded ANO solutions are the Z-strings,
which have been shown to be stable for
$\sin^2\theta_{\rm W}\gtrsim0.9$ \cite{instab}.

Recently, a new family of current carrying, ``twisted'' string solutions have been discovered
in the SU(2) symmetric semilocal model \cite{FRV}. 
These solutions are characterized by a relative phase
difference, $\exp(i\omega z)$, between the two components of the scalar field, 
where $z$ is the coordinate along the string
and $\omega$ is the twist.
Twisted strings exist only for $\beta>1$ where the embedded ANO solutions are unstable.
Quite remarkably 
the energy of twisted strings is lower than that of the
embedded ones (as a matter of fact the energy per unit length of twisted strings is a monotonously
decreasing function of the twist $\omega$).
The instability of the embedded ANO vortices for $\beta>1$ can be seen to correspond to a bifurcation
with twisted strings.
Clearly the problem of stability of twisted strings is an important one. 

A linear stability analysis of twisted strings has been presented by Ref.\ \cite{volkov}, 
and a family of unstable modes has been found. More precisely 
for all values of the twist $\omega$, the lowest eigenvalue belongs to a nonzero value of the wave number $k$ of the eigenmode
in the longitudinal direction. In particular there are no homogeneous ($z$-independent) negative energy eigenmodes.
Noting the analogy between the harmonic $z$-dependence of the instability mode and the Plateau--Rayleigh instability
in hydrodynamics (i.e.\ the fragmentation of a fluid stream into droplets), it has been argued in Ref.\ \cite{volkov} 
that the unstable eigenmode of twisted strings signals its breakup into small droplets. 

In this paper, we present a detailed analysis of the stability of twisted vortices. Working in a different gauge
than Ref.\ \cite{volkov} our work provides an independent check of the results of Ref.\ \cite{volkov}.
Both Ref.\ \cite{volkov} and the present
paper apply linearization analysis. In the present paper, following Ref.\ \cite{Goodband}, we work
in the background field gauge.
By solving the small fluctuation equations numerically for a large range of the parameters of the problem
we confirm the numerical results of Ref.\ \cite{volkov}, 
although we also find some small discrepancies which do not effect, however, the validity of the numerical results.
We also find that twisted strings possess a family of unstable modes, parametrized by their
wave number in the $z$ direction, $k\in[-k_{\rm m},k_{\rm m}]$.
These results are in agreement with those of Ref.\ \cite{volkov}.
In addition to a full fledged numerical approach, we have also developed a semi-analytical description
of the instability eigenmodes of twisted vortices for values of the twist
near the bifurcation point. In our view this sheds some light on the nature of the instability of twisted strings.
The description of the bifurcation also makes it possible to obtain the eigenvalues and eigenmodes of embedded strings 
as a deformation of those of twisted strings (with the deformation parameter being a function of the difference of the twist $\omega$
and the bifurcation point $\omega_{\rm b}$).
We note that twisted strings with periodic boundary conditions are also unstable in the semilocal
model. 

Clearly a linear stability analysis is not sufficient to draw definite conclusions on the issue of the final state 
of twisted strings.
There are nevertheless two important properties of the instabilities which can form the basis of some speculations on the 
long-time dynamics of
twisted vortex instability. Firstly, not far from the bifurcation with the ANO vortices, the unstable modes of twisted vortices
and those of the embedded
ANO ones are very similar. The field component, which dominates the eigenmodes close to the one corresponding to the largest negative
eigenvalue is nearly homogeneous. Secondly, independently of the
absence of $z$-independent instability eigenmodes for twisted strings
the {\sl $z$-independent} instability mode of the embedded ANO vortices is
still {\sl an energy lowering} perturbation. In such a case, a sufficiently general initial perturbation
(like a lump on the string, see Subsection \ref{sec:prop})
overlaps with the eigenmode corresponding to the most rapidly growing ones, those being almost $z$-independent.
Therefore we predict that twisted strings whose twist is close to the value at the bifurcation point
start to expand nearly homogeneously very similarly to the instability of embedded ANO vortices.
For twisted strings whose twist is is far from the bifurcation value this expansion may become local,
in the sense that expanded lumps may form on the strings.
The long-time dynamics, however, is expected to be rather different for twisted strings as compared to the
embedded ANO ones.
This difference is due to the global current flowing in twisted strings. 
We have found that to first order in perturbation theory the current
remains localized to the string.
Unless the current is completely carried away by radiative effects which come from higher orders in perturbation theory,
in contrast to the expansion of the embedded ANO strings \cite{akpv}, the expansion of twisted 
ones cannot go on indefinitely. 
It is also conceivable that the instability of static (or stationary) twisted strings signals 
that there are time dependent (oscillating) breather-like states.
We also present some arguments, that twisted strings are unlikely to break up
into small droplet-like configurations.

The outline of the paper is as follows: in Section \ref{sec:vort} we recapitulate the main characteristics of
the $SU(2)$ semilocal theory and its vortex solutions. In Subsection \ref{sec:BGbif} we give a
semi-analytical description of the bifurcation of twisted vortices with the embedded ANO vortex.
In Section \ref{sec:lin} we turn our attention to the stability problem of the twisted vortex solutions, and present the
numerical results. In Subsections \ref{sec:pertbif} and \ref{sec:perteig} the behavior of the linearized equations
near the bifurcation is studied.
In Subsection \ref{sec:prop} the properties of the eigenmodes are analyzed, and the possible
scenarios for the dynamics of the instability are presented.
The technical details of the calculations of Section \ref{sec:lin} are relegated to
Appendix \ref{app:perteqns}. In Appendix \ref{sec:local} a summary of the perturbations of embedded ANO vortices
is given.

\section{The $SU(2)$ semilocal theory and its vortices}\label{sec:vort}
\subsection{The $SU(2)$ semilocal theory}

The $4$ dimensional (4D) action of the SU(2) semilocal theory
can be transformed by suitable rescalings,
rendering the rescaled fields and coordinates to be dimensionless, and setting the charge of the scalar particle to unity,
to the form:
\be\label{4D-action}
S=\int\! d^4x\,\left\{-\frac14\,F_{\mu\nu}F^{\mu\nu}
+(D_\mu\Phi)^\dagger D^\mu\Phi-\frac{\beta}{2}\,
(\Phi^\dagger\Phi-1)^2\right\}\,.
\end{equation}
Where $F_{\mu\nu}=\partial_\mu A_\nu-\partial_\nu A_\mu$,
$D_\mu\Phi=\partial_\mu\Phi-iA_\mu\Phi$, $\Phi^T=(\phi_1,\phi_2)$.
The signature of the flat Minkowskian metric used here is $(+,-,-,-)$.
Because of the spontaneous breaking of the U(1) gauge symmetry
the physical spectrum contains two massive particles,
a scalar and a vector whose mass ratio is given by
$\beta=m_{\rm s}^2/m_{\rm v}^2$.
The fields transform under the U(1) gauge symmetry as
\be\label{gauge}
A_\mu\to A_\mu+\partial_\mu\Lambda(x)\,,\qquad
\Phi\to e^{i\Lambda(x)}\Phi\,,
\end{equation}
while the complex doublet, $\Phi^T=(\phi_1,\phi_2)$,
transforms as the fundamental representation of the global SU(2) symmetry.
The Euler-Lagrange equations following from the
action \eqref{4D-action} can be written as:
\begin{subequations}\label{4d-eqs}
\begin{align}
\pa^\rho F_{\rho\mu}& =i\{(D_\mu\Phi)^\dagger\Phi
-\Phi^\dagger D_\mu\Phi\}\,, \\
D_\rho D^\rho\Phi& = \beta(1-|\Phi|^2)\Phi\,,
\end{align}
\end{subequations}
where $|\Phi|^2=\Phi^\dagger\Phi=|\phi_1|^2+|\phi_2|^2$.
Any solution of the Abelian-Higgs model with a single scalar field,
$(A_\mu\,,\phi_1)$, can be embedded into the semilocal model
simply by putting $(A_\mu\,,\Phi:=\phi_1\Phi_0)$ where
$\Phi_0$ is a constant SU(2) doublet of unit norm, $\Phi_0^\dagger\Phi_0=1$.

The SU(2) symmetric semilocal theory has the following conserved Noether currents
\begin{equation}
  \label{eq:Noether}
  j_\mu^{\hat{a}} = -i\left( (D_\mu\phi_a)^* T^{\hat{a}}_{ab}\phi_b - \phi_a^*T^{\hat{a}}_{ab}D_\mu\phi_b\right)\,,
\end{equation}
where $\hat{a}=(0,a)$ and $T^{\hat{a}}=(1,\tau^a)$ with $\tau^a$ being the Pauli matrices.
The currents $j_\mu^{a}$ generate the global SU(2) while $j_\mu^0$ the local U(1) symmetry.

We shall also consider the special theory corresponding to the $\beta\to\infty$ limit.
In this limiting case, the scalar fields are
constrained by $\Phi^\dagger\Phi\equiv1$, and the limit theory is nothing
but a gauged $\mathbf{CP}^1$-model. The constraint can be taken
into account by replacing the scalar potential in the Lagrangian
of Eq.\ \eqref{4d-eqs} by a Lagrange multiplier term:
\begin{equation}
  \label{eq:BVlag}
  \lag_\infty=-\frac{1}{4}F_{\mu\nu} F^{\mu\nu} + (D_\mu \phi_a)^* D^{\mu}\phi_a -\lambda(|\phi|^2-1).
\end{equation}
The equations of motion in this case are very similar to the regular case,
\begin{equation}\label{eq:FRVMEBV}
\begin{aligned}
  D_\mu D^\mu \phi_a &= -\lambda\phi_a \\
  \partial^\mu F_{\mu\nu}\, &= -i(\phi_c^* D_\nu\phi_c -\phi_c (D_\nu\phi_c)^*),
\end{aligned}
\end{equation}
with
\begin{equation}
  \label{eq:LagMult}
  \lambda = \phi_a^* D_\mu D^\mu \phi_a = \phi_a (D_\mu D^\mu \phi_a)^* \,.
\end{equation}

\subsection{Vortex solutions}
To obtain twisted vortex solutions of the theory (\ref{4D-action}), one needs to write up the most
general axially symmetric Ansatz \cite{FRV}. Here, ``axially symmetric'' is meant in the general
sense, that rotations around the symmetry axis, or translations in its direction can be compensated
by a suitable gauge transformation \cite{FM}.

Let us choose the $z=x_3$ coordinate along the symmetry axis. Choosing
a suitable reference frame the configuration can be rendered static, which choice shall
be assumed from now on. The axially symmetric configuration has rotational symmetry in
the $(x_1\,,x_2)$ plane, and is then described by the following general Ansatz:
\begin{equation}
  \label{eq:FRVAns}
\begin{aligned}
  \phi_1(r,\vartheta,z)\, &= f_1(r) e^{i n\vartheta}, \\
  \phi_2(r,\vartheta,z)\,  &= f_2(r) e^{i m\vartheta}e^{i\omega z}, \\
  A_\vartheta(r,\vartheta,z) &= n a(r),\\
  A_3(r,\vartheta,z) &= \omega a_3(r),
\end{aligned}
\end{equation}
with $A_0=0$, $A_r=0$. The parameter $\omega$, which describes the $z$ dependence of the relative
phase of the two scalar field components, is the twist.

In what follows we shall consider the simplest and most
important class of configurations with $m=0$ and with $n=1$, but of course
our analysis can be easily extended for the general case.
The vortex profile functions in (\ref{eq:FRVAns}) obey the equations
\begin{equation}
  \label{eq:FRVprofE}
  \begin{aligned}
    \frac{1}{r}(r a_3')' &= 2 a_3 |f|^2-2 f_2^2,\\
    r\left(\frac{a'}{r}\right)' &= 2 f_1^2(a-1) + 2 f_2^2 a \,,\\
    \frac{1}{r}(r f_1')' &= f_1\left[ \frac{(1-a)^2 n^2}{r^2} + \omega^2 a_3^2 -\beta(1-|f|^2)\right],\\
    \frac{1}{r}(r f_2')' &= f_2\left[ \frac{(n a)^2}{r^2} + \omega^2 (1-a_3)^2 -\beta(1-|f|^2)\right].
  \end{aligned}
\end{equation}
Regularity at the origin $r=0$ is ensured by the boundary conditions 
$a\to 0$, $f_1 \to 0$, $f_2\to c$, $a_3\to c'$, where $c$, $c'$ are constants,
while for $r\to\infty$  we demand  $a\to 1$, $f_1 \to 1$ $f_2,a_3\to 0$.

The embedded Abrikosov-Nielsen-Olesen (ANO) solutions
correspond to $f_2=a_3=0$ in Eqns.\ (\ref{eq:FRVprofE}). Besides these, new classes of vortex solutions have been discovered.
First, Ref.\ \cite{hin} found a one-parameter family of solutions for $\beta=1$ with $f_2\ne0$ and $a_3=0$.
The parameter of the vortices in this family of solutions can be identified with their
width (denoting the parameter with $\xi$, the radial decay of the magnetic field is $1-(r/\xi)^{-2}$).
These solutions satisfy  a Bogomol'nyi
type energy bound \cite{vac-ach}, therefore their energy is degenerate, and they obey simpler, first order field
equations.

Another one-parameter class of solutions has been discovered in \cite{FRV} for $\beta > 1$ with both $f_2$ and $a_3\ne 0$.
These solutions are parameterized by the value of their twist, $0<\omega<\omega_{\rm b}(\beta)$. At $\omega=\omega_{\rm b}$
these twisted vortices bifurcate with the embedded ANO vortex (for some numerical values of $\omega_{\rm b}(\beta)$ for
a range of $\beta$
see Ref.\ \cite{FRV}). There is a similarity between this family of vortices and the one
of Ref.\ \cite{hin}.
The parameter $\omega$ also characterizes the width of vortices: a lower value of $\omega$ corresponds
to a more diluted vortex with a slower radial decay of the magnetic field, and a higher value of the scalar condensate at the origin.
(However, one should note, that the radial decay of these solutions is different from that of the $\beta=1$ ones, as here,
$f_2$ decays with $\exp(-\omega r)$.)

An important property of twisted vortices is that there is a global current flowing along them. In the coordinate system
fixed by the Ansatz (\ref{eq:FRVAns}), the current is the third isospin component of the semilocal current
(\ref{eq:Noether}) in the $z$ direction. The total current flowing along the string can thus be written as
\begin{equation}
  \label{eq:curr2}
  I_3 = \int r\d r\d \vartheta j_3^3 \,.
\end{equation}
As the twist $\omega$ decreases, the current increases and the twisted vortex becomes more and
more spread out \cite{FRV}. It appears that the limit $\omega\to0$ is somewhat
singular, nevertheless in this limit twisted vortices converge pointwise to a configuration
which is precisely the large width limit of the $\beta=1$ semilocal vortices of Ref.\ \cite{hin}.

Another important property of twisted vortices is that they have lower energy than
the embedded ANO solutions. This indicates that such twisted vortices may be preferred by physical processes over
the embedded ANO vortices.
In fact, for fixed $\beta$, the energy monotonously decreases as $\omega$ decreases,
i.e.\ as we go further away from the bifurcation \cite{FRV}.

In the $\beta\to\infty$ limiting theory $f_1$ and $f_2$ are related by the constraint
$f_1^2+f_2^2\equiv1$. A convenient way of parameterizing them is
\begin{equation}
  \label{eq:BVans}
  f_1 = \sin\theta\,,\qquad f_2 = \cos\theta\,,
\end{equation}
which yields the profile equations
\begin{equation}
  \label{eq:BVprofE}
  \begin{aligned}
    \frac{1}{r}\left( r a_3'\right)   &= 2\left[a_3-\cos^2\theta\right],\\
    r\left(\frac{a'}{r}\right)'       &= 2(a-\sin^2\theta)\,,\\
    \frac{1}{r}\left(r\theta'\right)' &= \frac{1}{2}\left[(2a_3-1)\omega^2-
    \frac{n^2}{r^2}(2a-1)\right]\sin(2\theta)\,.
  \end{aligned}
\end{equation}
In this case the embedded ANO solution has infinite energy,
whereas the twisted vortices are of finite energy and they exist for all values of $\omega$ \cite{FRV}.

\subsection{Bifurcation of the embedded ANO solution}\label{sec:BGbif}
It is by now well known \cite{hin} that the embedded ANO vortices are unstable to small perturbations of the $f_2$ variable.
This instability corresponds to the bifurcation of the ANO vortices with the twisted vortex solutions \cite{FRV}.
The systematic expansion of a twisted vortex near the bifurcation point
can be then written as:
\begin{equation}
  \label{eq:bifureps}
  \begin{aligned}
    f_1 &= f_1^{(0)} +\epsilon^2 f_1^{(2)} + \ldots \\
    f_2 &= \epsilon f_2^{(1)}+ \epsilon^2 f_2^{(2)}+\ldots \\
    a   &=  a^{(0)}  + \epsilon^2 a^{(2)} + \ldots \\
    a_3 &=  \epsilon^2 a_3^{(2)} + \ldots \\
    \omega &= \omega_{\rm b} +\epsilon\omega_1+ \epsilon^2 \omega_2\,\,\, + \ldots
  \end{aligned}
\end{equation}
where $a^{(0)}\,,f_1^{(0)}$ denotes the ANO vortex, whose equations can be read off
from equations (\ref{eq:FRVprofE}) by putting $f_2=a_3=0$. In the above expansion, we have omitted some terms linear in $\epsilon$.
When solving the equations of the field components, these turn out to be zero (i.e.\ their equation
is a homogeneous linear equation). As we shall see the only first order term is $f_2^{(1)}$. This is physically plausible too:
the bifurcation is parameterized by the growing of the condensate $f_2$.

\begin{figure}[!ht]
\noindent\hfil \includegraphics[scale=.5,angle=-90]{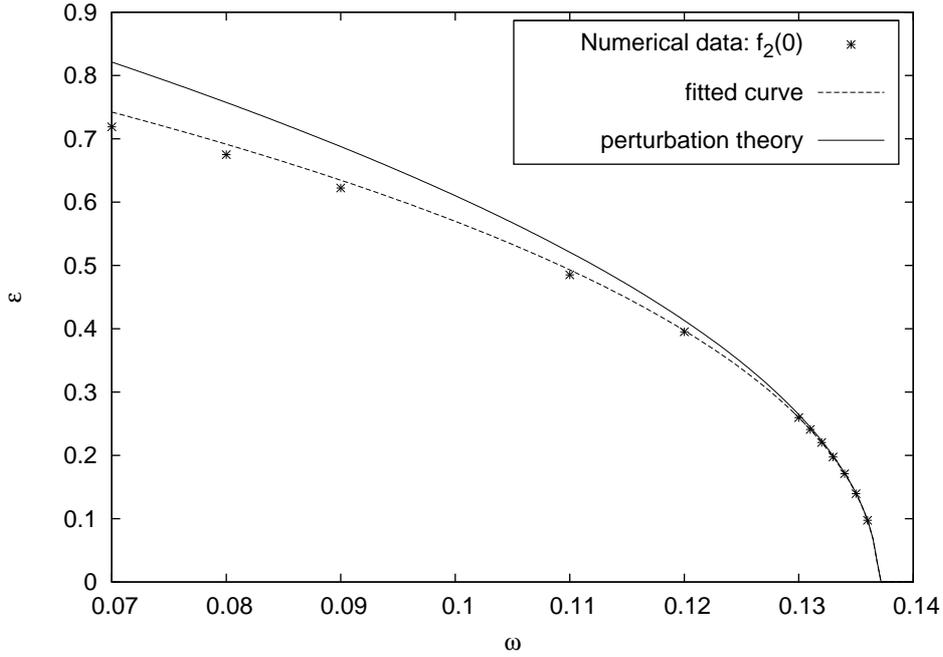}
\caption{The $\omega$--$\epsilon$ dependence, $\beta=1.25$}
\label{fig:omeps125}
\end{figure}

\begin{figure}[!ht]
\noindent\hfil\includegraphics[scale=.5,angle=-90]{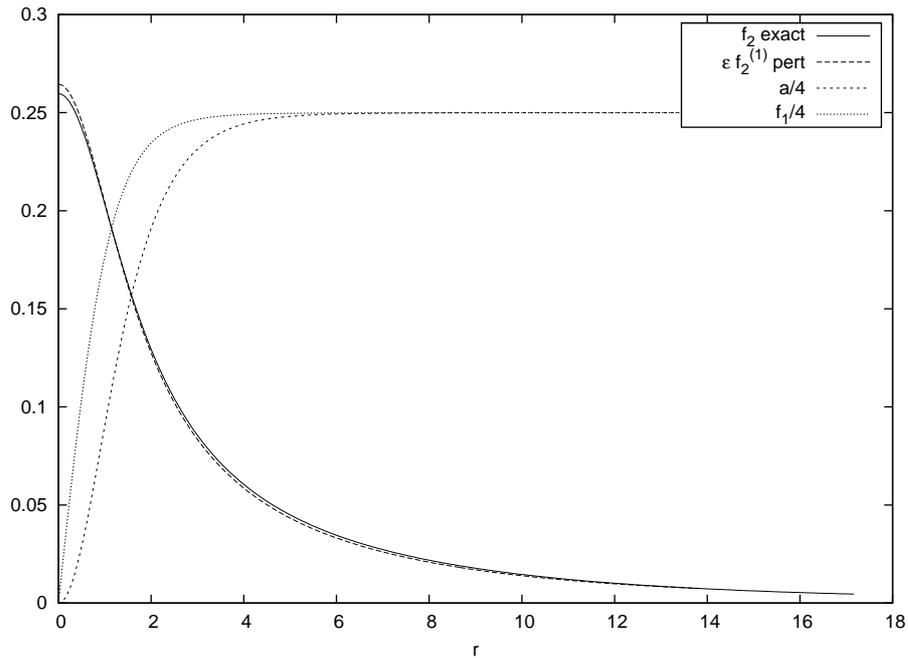}
\caption{The twisted vortex profile function $f_2$ for $\beta=1.25$, $\omega=0.13$. For comparison, we have
also plotted $f_1$ and $a$ in this figure (note the scaling by $1/4$ of $f_1$ and $a$).}
\label{fig:twistedf2}
\end{figure}

\begin{figure}[!ht]
\noindent\hfil\includegraphics[scale=.5,angle=-90]{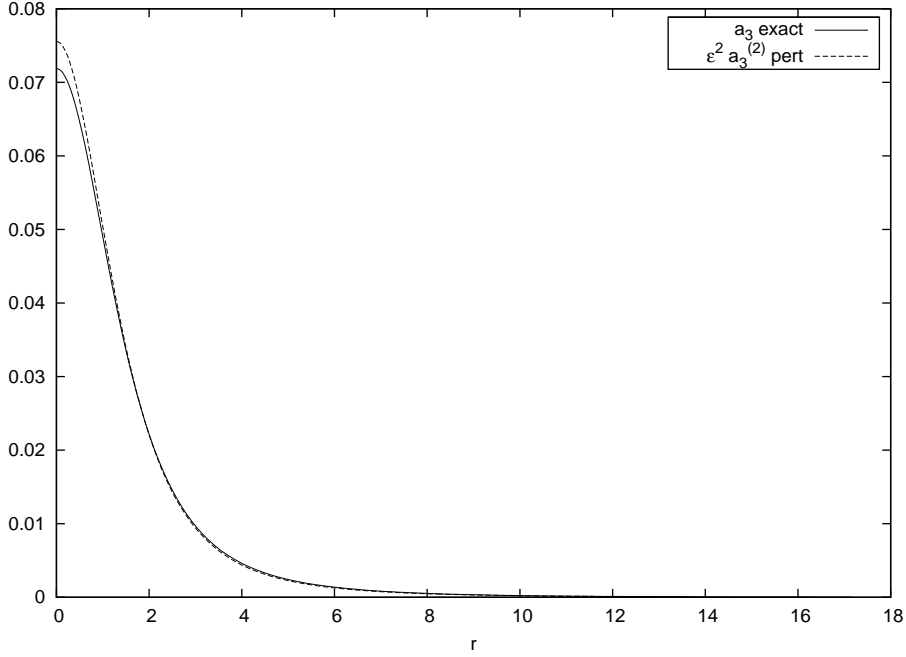}
\caption{The twisted vortex profile function $a_3$ for $\beta=1.25$, $\omega=0.13$}
\label{fig:twisteda3}
\end{figure}

The equations for the field components can be obtained easily by substituting the
above expansion into the vortex profile equations (\ref{eq:FRVprofE}).
To first order in the $\epsilon$ expansion the only non-trivial equation
determines the second component of the scalar field,
\begin{equation}\label{eq:bifur-f2}
(D_2^{(0)}+ \omega_{\rm b}^2)f_2^{(1)}
  := -\frac{1}{r}\left(r {f_2^{(1)}}'\right)' + \left[ \frac{(na^{(0)})^2}{r^2} -
 \beta \left(1-(f_1^{(0)})^2\right)+ \omega_{\rm b}^2\right] f_2^{(1)}=0\,.
\end{equation}
(The operator $D_2^{(0)}$  should not be confused with the $\mu=2$
component of covariant derivative $D_\mu$.)
In second order the equation for $f_2^{(2)}$ can be written as
\begin{equation}\label{eq:f22eq}
  (D_2^{(0)}+\omega_{\rm b}^2) f_2^{(2)} = - 2 \omega_{\rm b} \omega_1 f_2^{(1)} \,.
\end{equation}
Eq.\ (\ref{eq:f22eq}) contains only a resonance term on the right hand side, therefore
the solution of this equation is $f_2^{(2)} = 0$
and the removal of the resonance term yields $\omega_1=0$.
\begin{equation}\label{eq:eps3}
  \begin{array}{c}
    \left(r {a_3^{(2)}}'\right)'/r - 2 (f_1^{(0)})^2 a_3^{(2)} = -2 (f_2^{(1)})^2 \,,\\
    \left(r {f_1^{(2)}}'\right)'/r - \left[ \displaystyle\frac{n^2(1- a^{(0)})^2}{r^2}-
    \beta(1-3(f_1^{(0)})^2)\right]f_1^{(2)}
    + f_1^{(0)} \displaystyle\frac{2n^2(1-a^{(0)})}{r^2} a^{(2)} =  \beta f_1^{(0)} (f_2^{(1)})^2 \,,\\
    r\left(\displaystyle\frac{{a^{(2)}}'}{r}\right)' - 2 (f_1^{(0)})^2 a^{(2)} -
    4 (a^{(0)}-1)f_1^{(0)}f_1^{(2)} = 2a^{(0)}(f_2^{(1)})^2\,.
  \end{array}
\end{equation}
The absence of the resonance term in Eq.\ (\ref{eq:eps3}) gives
\begin{equation}
  \label{eq:omega2}
  \omega_2 = -\frac{1}{2\omega_{\rm b}\|f_2^{(1)}\|}\left(f_2^{(1)},\frac{2n a^{(0)}}{r^2} n a^{(2)}f_2^{(1)}-2\omega_{\rm b}^2a_3^{(2)}f_2^{(1)}+2\beta
  f_1^{(0)}f_1^{(2)}f_2^{(1)}+\beta (f_2^{(1)})^3\right) \,,
\end{equation}
where the scalar product and the norm is defined as
\begin{equation}
  \label{eq:scalarprod}
  (f,g):=\int_0^\infty r \d r f g \,, \qquad \text{and} \qquad \|f\|^2 := (f,f)\,.
\end{equation}
Identifying  $f_2(0)$ with the perturbation parameter, $\epsilon$, at the bifurcation
point, we obtain that
\begin{equation}
\epsilon = \sqrt{\frac{1}{\omega_2}(\omega-\omega_{\rm b})} + \dots\,.
\end{equation}
We have compared the the results of the above perturbative bifurcation analysis,
for  $f_2(0)$ as a function of $\omega$
with numerical results on
Figure \ref{fig:omeps125}. 
A fitted curve, of the form
$\omega_2^{-1/2}(\omega_{\rm b}-\omega)^{1/2}+\epsilon_1 (\omega_{\rm b}-\omega)$
, with $\omega_2$ and $\epsilon_1$ being the parameters fitted, as suggested by
perturbation theory, is also shown. See also Table \ref{tab:pert}.
Another nice comparison of the twisted and ANO vortices can be seen on Figures
\ref{fig:twistedf2}-\ref{fig:twisteda3}. We have drawn no figures of $f_1$ and $a$. Their perturbation series
starts with a zeroth order ANO term, and the next correction is of order $\epsilon^2$, therefore, the difference between
the ANO background, the second order perturbative result and the exact twisted vortex profile function is quite small.

As we can see from both the comparison of the $\epsilon$ -- $\omega$ dependence and
the vortex profile functions, the above perturbative method gives a good approximation of twisted
vortices, even relatively far from the bifurcation, i.e.\ for $\epsilon=0.2\dots 0.3$. In Subsections \ref{sec:pertbif}  
and \ref{sec:perteig} we will use this perturbatively obtained vortex background to explore the relation between the
instabilities of the embedded ANO vortices and those of the twisted ones.

\section{Linear stability analysis}\label{sec:lin}
\subsection{Perturbation analysis}\label{sec:linpert}

Here we present a stability analysis of the twisted vortex solution based on the linearization of the field equations about
the vortex. Let $\phi_a = \phi_{a,\text{bg}} + \epsilon\delta\phi_a$ and $A_\mu=A_{\mu,\text{bg}}+\epsilon\delta A_\mu$,
with $\phi_{a,\text{bg}}$ and $A_{\mu,\text{bg}}$ being the background solution. The index ``bg'' will be dropped in what follows;
the total perturbed fields will not appear later on. Here, we will summarize the main properties and the method of solution of the
linear equations governing $\delta\phi_a$ and $\delta A_\mu$.  Following the footsteps of Ref.\ \cite{Goodband}
we will use the background field gauge. The main differences between the present case and that of the ANO vortices is that here
there are more field components and the $z$-dependence does not decouple, which makes the problem computationally more demanding.
The main advantage of the background field gauge is the simplicity of the resulting equations. Its main drawback is, that, unlike
the method used by Ref.\ \cite{volkov}, it does not fix the gauge completely, which makes an analysis of ghost modes necessary.

The perturbation equations can be derived from a linearization of the field equations. The resulting equations, taking into account
that the background solution satisfies the Lorentz gauge condition $\partial^\mu A_{\mu}=0$, can be cast into the form
\begin{equation}
  \label{eq:FRVlag2a}
  D \begin{pmatrix}
    \delta\phi_a \\ \delta\phi^*_a \\ \delta A^\mu
\end{pmatrix} = 0,
\end{equation}
where the operator $D$ is
\begin{equation}
  \label{eq:FRVlag2}
\begin{pmatrix}
D_1 + 2i\delta_{ab}A^\mu\partial_\mu -\beta \phi_b\phi^*_a & -\beta \phi_b \phi_a & 2i\partial_\mu\phi_b + 2 A_\mu\phi_b +i\phi_b\partial_\mu \\
-\beta \phi^*_b\phi^*_a & D_1 - 2i\delta_{ab}A^\mu\partial_\mu -\beta \phi^*_b\phi_a  & -2i\partial_\mu\phi^*_b + 2 A_\mu\phi^*_b -i\phi^*_b\partial_\mu \\
-i\partial_\nu \phi^*_a + 2A_\nu\phi^*_a + i\phi^*_a\partial_\nu & i\partial_\nu \phi_a + 2A_\nu\phi_a - i\phi_a\partial_\nu &
|\phi|^22g_{\mu\nu}-\partial_\mu\partial_\nu +g_{\mu\nu}\square
\end{pmatrix},
\end{equation}
with $D_1=-\delta_{ab}\square +A_\mu A^\mu\delta_{ab}-\beta (|\phi|^2-1)\delta_{ab}$ and $\square=\partial_\mu \partial^\mu$.

In the $\beta\to\infty$ limit, similarly to the regular case, the Lagrangian is expanded to second order in $\delta\phi_a$,
$\delta\phi_a^*$ and $\delta A_\mu$. Denoting the zeroth and first order terms of $D_{\mu}D^{\mu}\phi_a$ with $\chi_a$ and $\psi_a$, equation
(\ref{eq:FRVpertEig}) can be written as $\psi_a=0$. Now, the linearization of the Lagrange multiplier (\ref{eq:LagMult})
has to be added. The resulting equations of motion are
\begin{equation}
  \label{eq:BVpertEQ}
  \begin{aligned}
    \psi_a - (\lambda^{(0)}\delta\phi_a+\lambda^{(1)}\phi_a) &= 0, \\
    \psi_a^* - (\lambda^{(0)}\delta\phi_a^*+\lambda^{(1)}\phi_a^*) &= 0 ,
  \end{aligned}
\end{equation}
where
\begin{equation}
  \label{eq:LagMultP}
  \begin{aligned}
    \lambda^{(0)} &= \frac{1}{2}\left(\phi_a^*\chi_a+\phi_a\chi_a^*\right), \\
    \lambda^{(1)} &= \frac{1}{2}\left( \delta\phi_a^*\chi_a+\phi_a\psi_a+\delta\phi_a\chi_a^*+\phi_a\psi_a^*\right).
  \end{aligned}
\end{equation}

The perturbation equations are invariant under infinitesimal gauge transformations of the form
\begin{equation}
  \label{eq:FRVinfgauge}
  \begin{aligned}
  \delta \phi_a &\rightarrow \delta \phi_a + i\chi\phi_a,\\
  \delta A_\mu  &\rightarrow \delta A_\mu + \partial_\mu \chi.
  \end{aligned}
\end{equation}
This gauge freedom can be dealt with by adding a gauge-fixing term $\frac{1}{2}|F(A)|^2$ to the Lagrangian,
\begin{equation}
  \label{eq:FRVBKGRGauge}
  F(A) := \partial_\mu \delta A^\mu + i(\delta\phi^*_a \phi_a - \phi^*_a\delta\phi_a) =0.
\end{equation}
This helps one to get rid of the terms containing first order derivatives in $D$, yielding the perturbation matrix
\begin{equation}
  \label{eq:FRVlag2b}
\begin{pmatrix}
D_1 + 2i\delta_{ab}A^\mu\partial_\mu -(\beta+1) \phi_b\phi^*_a & -(\beta-1) \phi_b \phi_a & 2i\partial_\mu\phi_b + 2 A_\mu\phi_b \\
-(\beta-1) \phi^*_b\phi^*_a & D_1 - 2i\delta_{ab}A^\mu\partial_\mu -(\beta+1) \phi^*_b\phi_a  & -2i\partial_\mu\phi^*_b + 2 A_\mu\phi^*_b \\
-2i\partial_\nu \phi^*_a + 2A_\nu\phi^*_a & 2i\partial_\nu \phi_a + 2A_\nu\phi_a &
|\phi|^22g_{\mu\nu} +g_{\mu\nu}\square
\end{pmatrix},
\end{equation}
where $D_1=-\delta_{ab}\square +A_\mu A^\mu\delta_{ab}-\beta (|\phi|^2-1)\delta_{ab}$.

It can be easily verified that the equation of $\delta A_0$,
\begin{equation}
  \label{eq:FRVa0}
  \partial_\mu\partial^\mu \delta A_0 + 2 |\phi|^2\delta A_0 = 0 \,,
\end{equation}
decouples, just as in the case of the ANO vortex \cite{Goodband}.

The gauge condition (\ref{eq:FRVBKGRGauge}) does not fix the gauge completely, but still
allows gauge transformations whose generating function
satisfies the ghost mode equation
\begin{equation}\label{eq:FRVghost}
\partial_\mu\partial^\mu \chi + 2|\phi|^2\chi = 0 \,.
\end{equation}
The ghost modes cancel the $\delta A_0$ modes, whose Eq.\ (\ref{eq:FRVa0}) is identical to (\ref{eq:FRVghost}), and
a part of the spectrum of the gauge fixed operator (\ref{eq:FRVlag2b}). In the case of the embedded ANO vortices,
a straightforward analysis of ghost modes is possible, see Appendix \ref{sec:local}.

Let us now apply a Fourier transform to the perturbation
equations (\ref{eq:FRVlag2b}) in the
$t$ and $z$ variables. Introducing $\Psi =(\delta\phi_1,\delta\phi_1^*,\delta\phi_2,\delta\phi_2^*,
\delta A_i,\delta A_3, \delta A_0)^T$.
\begin{equation}
  \label{eq:Fourier}
  \Psi = \int \d k \d \Omega e^{i(\Omega t - k z)} \tilde\Psi.
\end{equation}
This way, the perturbation equations can be brought into the form
\begin{equation}
  \label{eq:FRV-fourier}
  \mathcal{M}  \tilde\Psi = \Omega^2 \tilde\Psi.
\end{equation}
See Appendix \ref{app:perteqns} for the details of the calculation and the matrix $\mathcal{M}$.
The angle variable $\vartheta$ can be separated
in a similar fashion, by the Fourier series expansion
\begin{equation}
  \label{eq:Fourtheta}
  \tilde\Psi = \sum_\ell \Psi_\ell e^{i\ell\vartheta}.
\end{equation}
This yields the desired eigenvalue problem
\begin{equation}
  \label{eq:FRVell}
  M_\ell  \Psi_\ell = \Omega^2 \Psi_\ell\,,
\end{equation}
where $M_\ell$ is an ordinary differential operator in the radial variable $r$ (see equation (\ref{eq:FRVpertEig}) in Appendix \ref{app:perteqns}).
A similar expansion can be done for the ghost modes.

\begin{figure}[!ht]
\noindent\hfil\includegraphics[scale=.3,angle=-90]{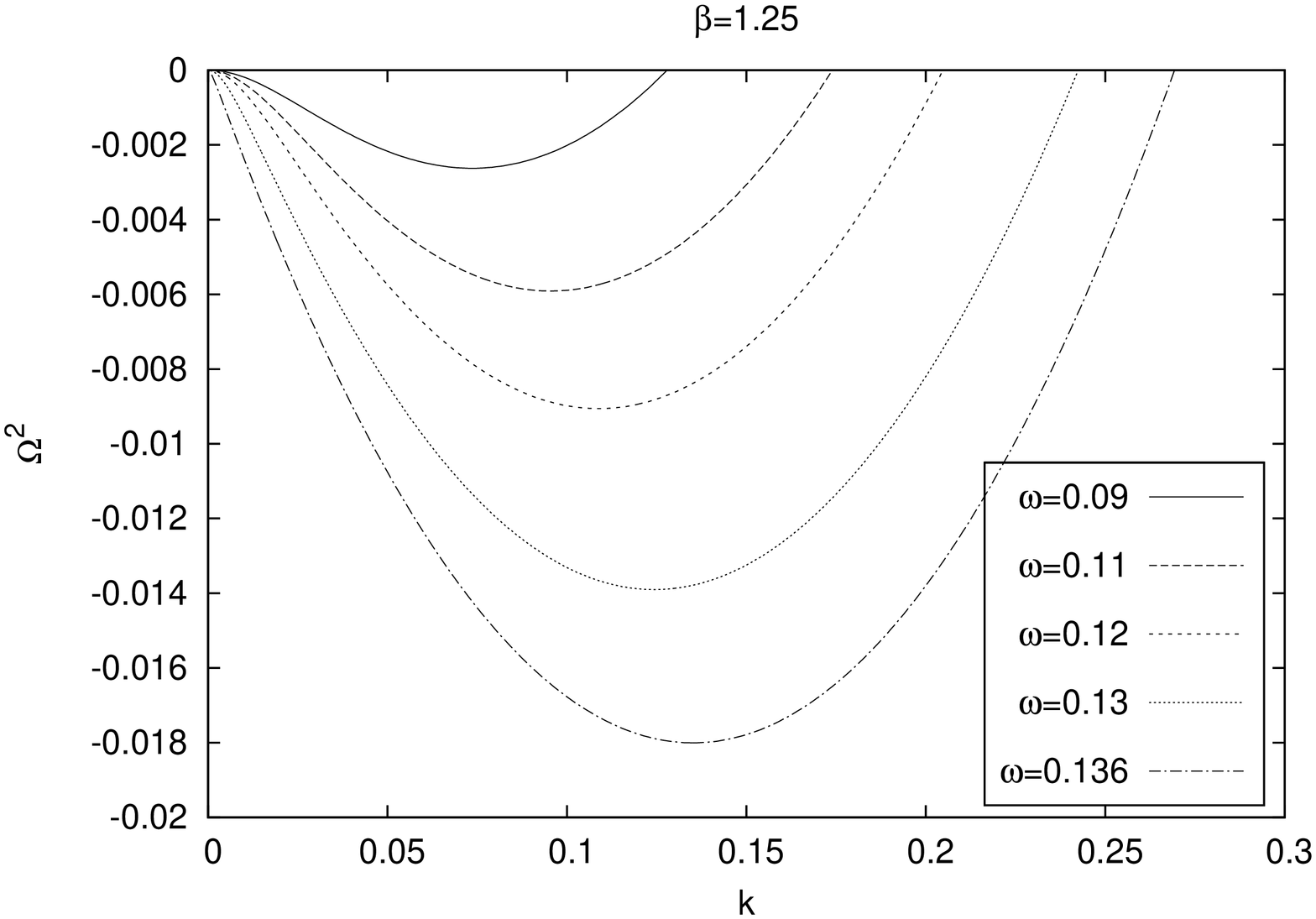} \includegraphics[scale=.3,angle=-90]{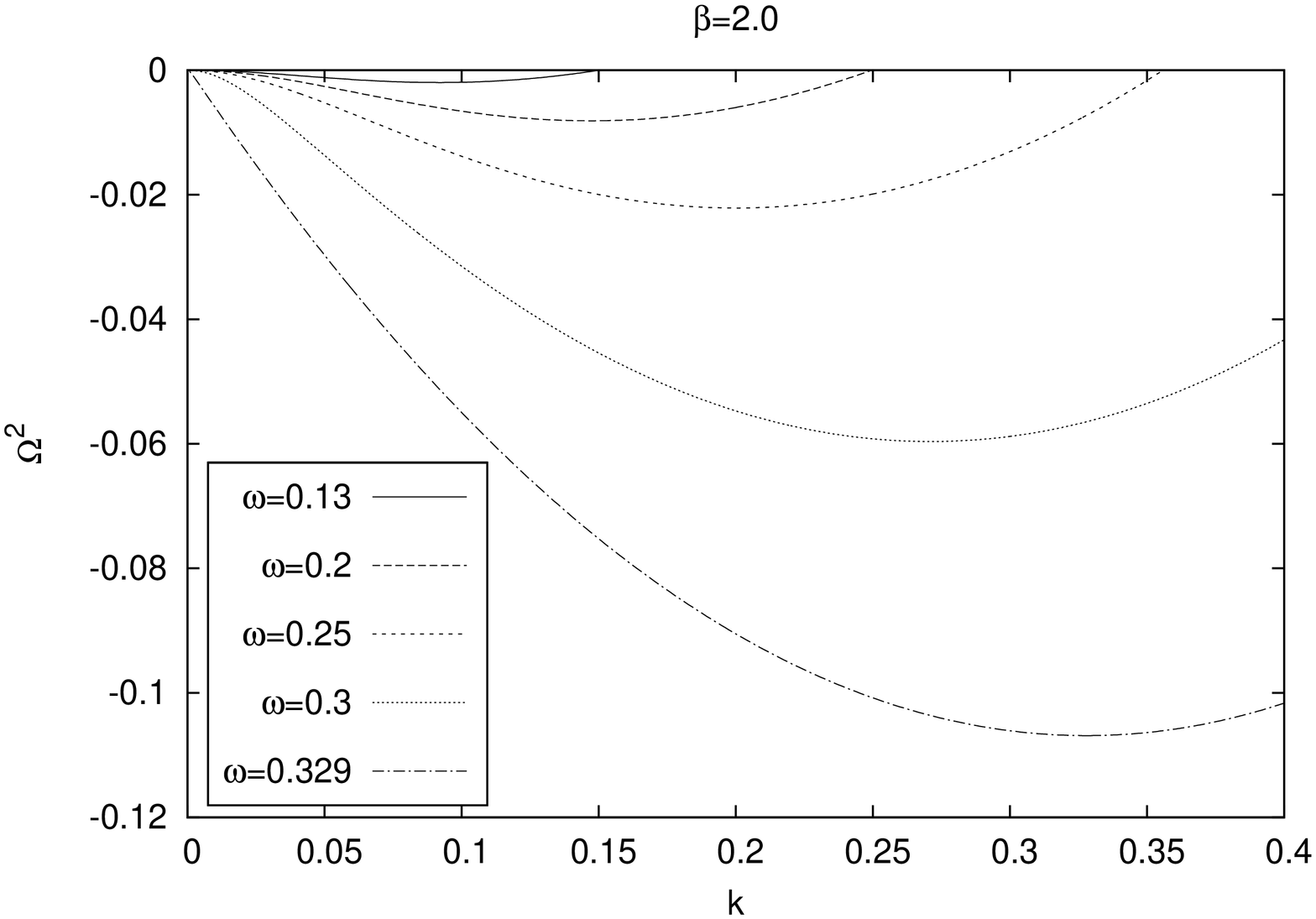}

\noindent\hfil\includegraphics[scale=.3,angle=-90]{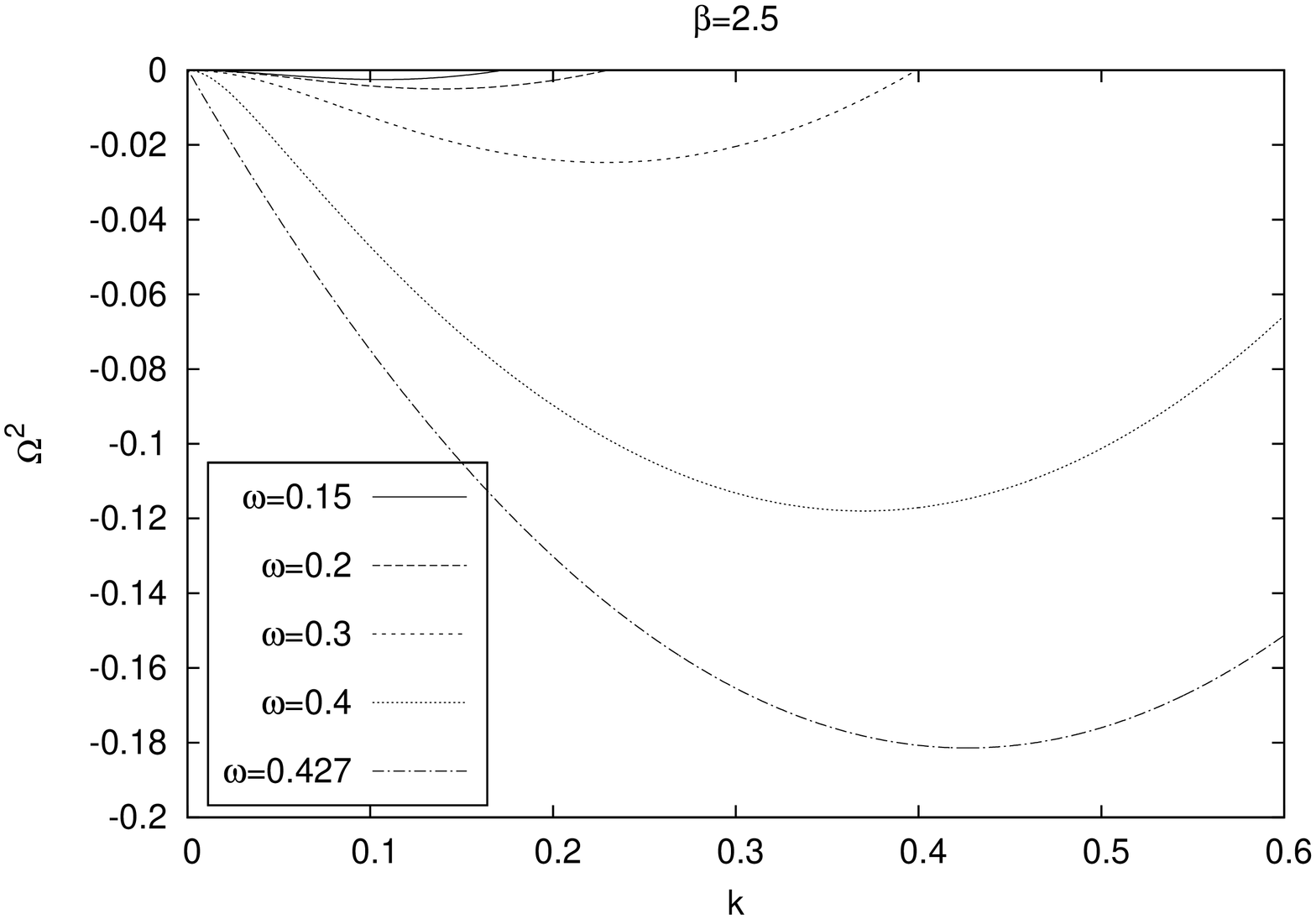} \includegraphics[scale=.3,angle=-90]{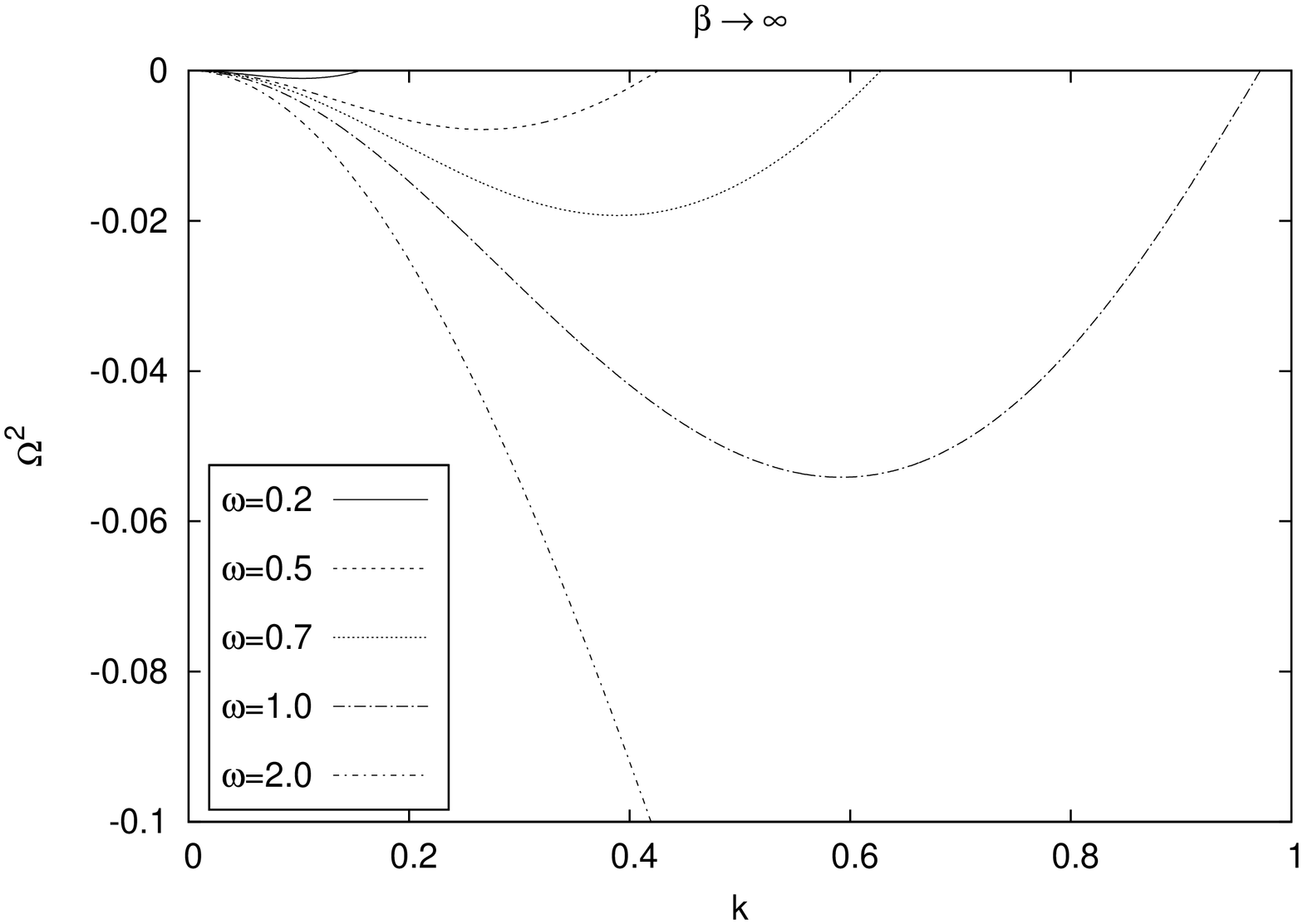}
\caption{The dispersion relation of the unstable mode}
\label{fig:disprel}
\end{figure}

The numerical method used for solving equations (\ref{eq:FRVell}) as well as the background equations (\ref{eq:FRVprofE}) was shooting to a fitting point.
The resulting eigenvalues as a function of $k$ can be seen on Figure \ref{fig:disprel} for $\beta=1.25$, $2$, $2.5$ and the limit $\beta\to\infty$.
There are two unstable modes,
one which is the deformation of the unstable mode in $s_2$ of the embedded ANO vortex, which has a negative eigenvalue for 
$0<k<k_{\rm m}$,
and another one, which is the deformation of the unstable mode in $s_2^*$ of the embedded ANO vortex and has a negative eigenvalue for
$0>k>-k_{\rm m}$.
These two modes are related by a transformation of $k\to-k$ and interchanging the fields $s_1,s_2,a_+$ and
the conjugate fields $s_1^*,s_2^*,a_-$. The numerical results exhibit this symmetry to a rather high accuracy.

\begin{figure}[!ht]
\noindent\hfil\includegraphics[scale=.3,angle=-90]{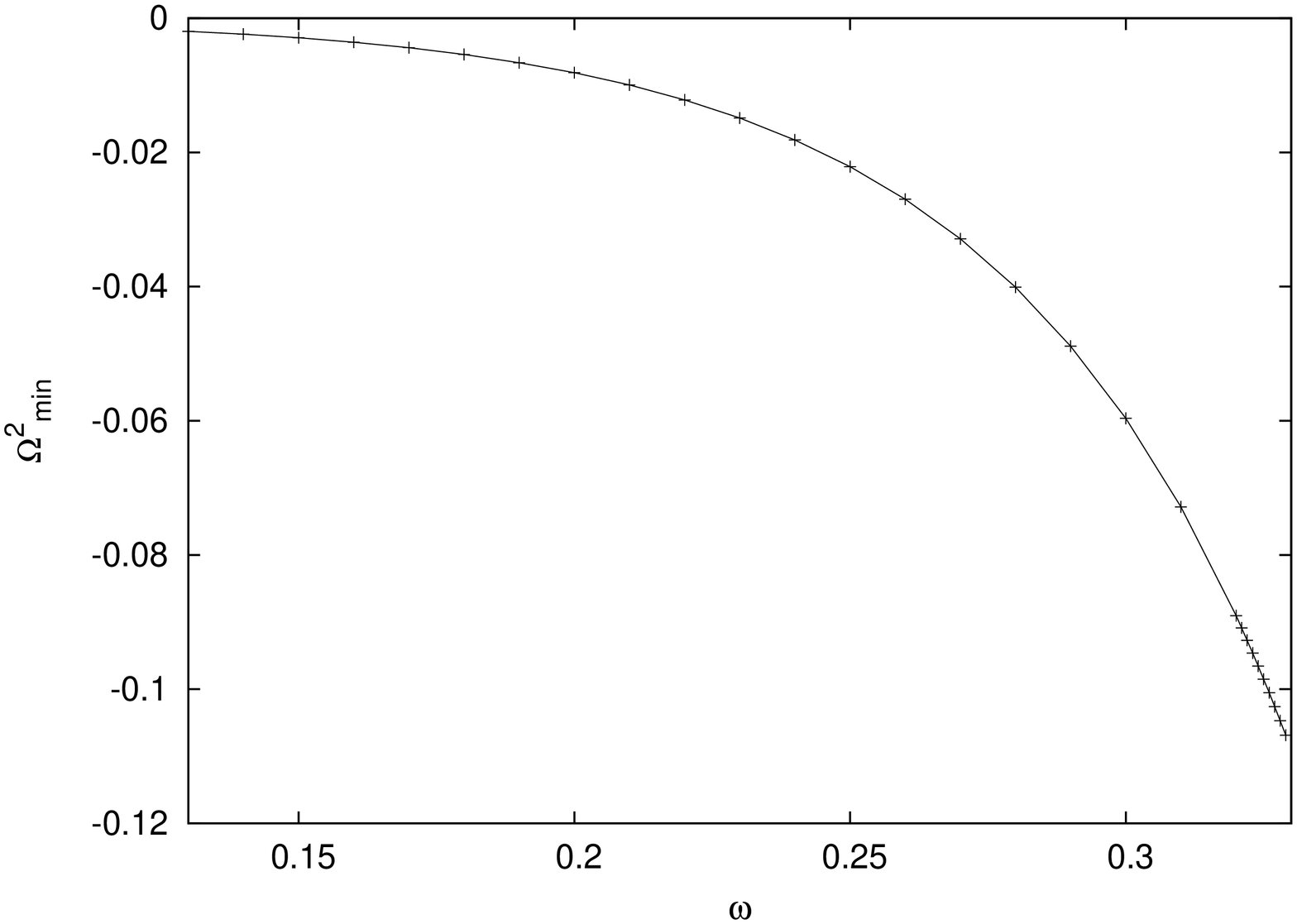} \includegraphics[scale=.3,angle=-90]{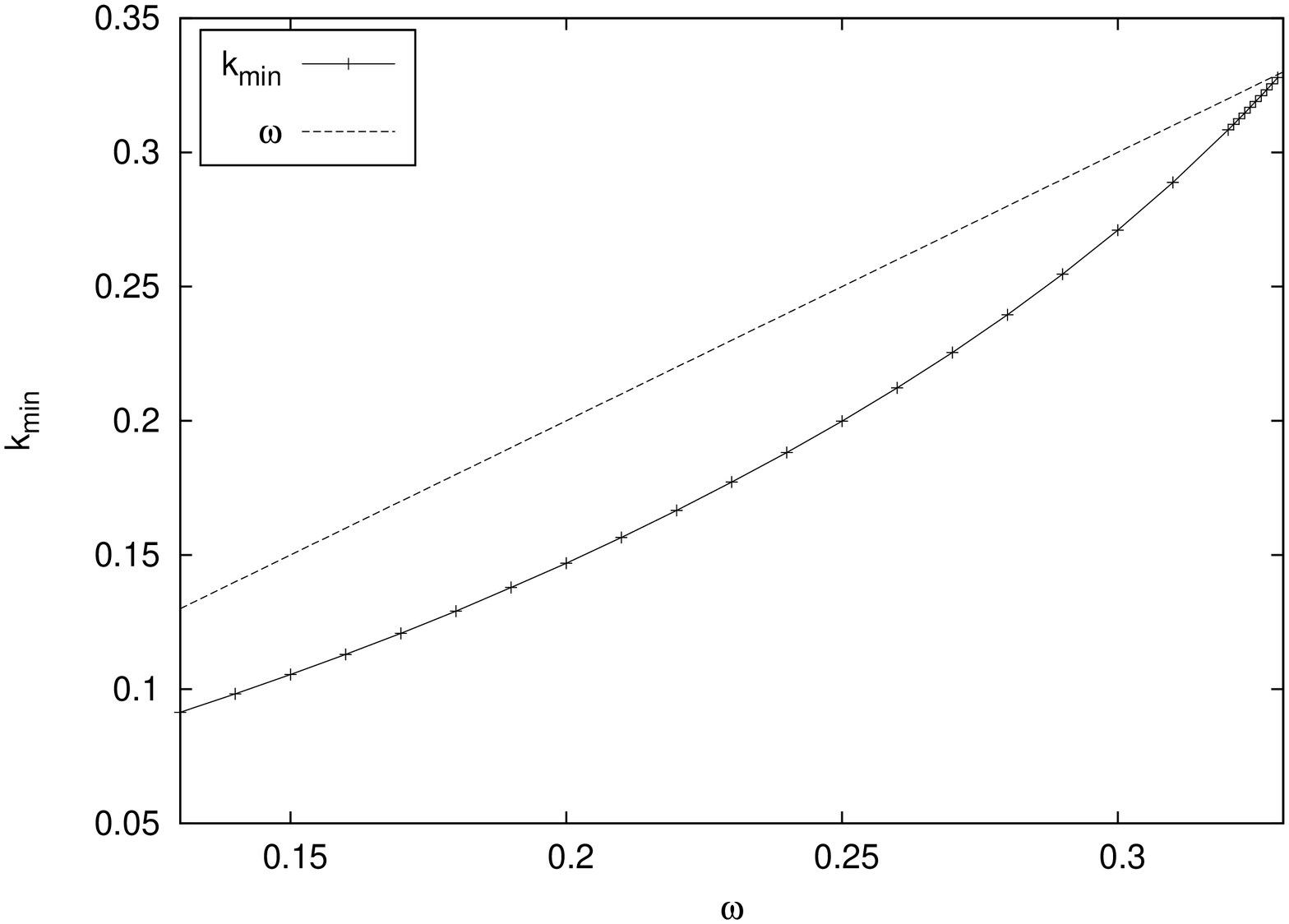}
\caption{Minima of $\Omega^2$ and its location for $\beta=2$ as a function of $\omega$}
\label{fig:omOm2}
\end{figure}

In Figure \ref{fig:disprel}, we have plotted the dispersion relation of the unstable modes for $0<k<k_{\rm m}$.
A closed form of this is not known, however, the approximation
\begin{equation}
  \label{eq:FRVpertdisp}
  \Omega^2 = \Omega_{\text{min}}^2 + \Omega_2^2 (k-k_{\text{min}})^2
\end{equation}
is empirically found to be very good.
Note that $k_{\text{min}}$, the wavenumber corresponding to the minimal eigenvalue, should not be confused with
$k_{\rm m}$ standing for the maximal value of the wave numbers of the unstable modes.
In Figure \ref{fig:omOm2}
one can see
$\Omega_{\text{min}}^2$ and $k_{\text{min}}$ as a function of $\omega$ for $\beta=1.25$. 
The same data, for $\beta=1.25$, $2$ and $2.5$ is summarized in Tables \ref{tab:pertdata125}, \ref{tab:pertdata2} and \ref{tab:pertdata25}.
In the case of the embedded ANO vortex, the form (\ref{eq:FRVpertdisp}) of the dispersion relation is exact.
In Ref.\ \cite{volkov} it is claimed that the minimum value is achieved at $k=\pm \omega$, but for smaller values of
$\omega$ we have found a considerable deviation from
$k_{\text{min}}=\pm\omega$ (see especially the curves corresponding to $\beta\to\infty$).

Close to the bifurcation, our numerical results agree with those of Ref.\ \cite{volkov}. However, for smaller values of $\omega$,
there is some discrepancy which we attribute to numerical errors.

\begin{table}[!ht]
\begin{center}
\begin{tabular}{|c||c|c|c|}
\hline
$\omega$ & $\Omega_{\text{min}}^2$ & $\Omega_2^2$ & $k_{\text{min}}$ \\
\hline\hline
0.07     & -0.00105               & 0.73927     & 0.05418    \\
0.08     & -0.00173               & 0.80044     & 0.06383    \\
0.09     & -0.00262               & 0.84997     & 0.07352    \\
0.11     & -0.00591               & 0.94158     & 0.09542    \\
0.12     & -0.00906               & 0.97588     & 0.10882    \\
0.13     & -0.01390               & 0.99548     & 0.12435    \\
0.136    & -0.01801               & 0.99984     & 0.13513    \\
\hline
\end{tabular}\end{center}
\caption{Twisted vortex perturbations: parameters of the dispersion relation for $\beta=1.25$}
\label{tab:pertdata125}
\end{table}

\begin{table}[!ht]
\begin{center}
\begin{tabular}{|c||c|c|c|}
\hline
$\omega$ & $\Omega_{\text{min}}^2$ & $\Omega_2^2$ & $k_{\text{min}}$ \\
\hline\hline
0.13     & -0.00197               & 0.53731     & 0.09140 \\
0.15     & -0.00293               & 0.57959     & 0.10545 \\
0.2      & -0.00812               & 0.73531     & 0.14689 \\
0.25     & -0.02213               & 0.88462     & 0.19989 \\
0.3      & -0.05964               & 0.97901     & 0.27098 \\
0.32     & -0.08905               & 0.99561     & 0.30837 \\
0.329    & -0.10687               & 0.99969     & 0.32785 \\
\hline
\end{tabular}\end{center}
\caption{Twisted vortex perturbations: parameters of the dispersion relation for $\beta=2$}
\label{tab:pertdata2}
\end{table}

\begin{table}[!ht]
\begin{center}
\begin{tabular}{|c||c|c|c|}
\hline
$\omega$ & $\Omega_{\text{min}}^2$ & $\Omega_2^2$ & $k_{\text{min}}$ \\
\hline\hline
0.14     & -0.00211               & 0.50713      & 0.09848 \\
0.15     & -0.00249               & 0.51283      & 0.10565 \\
0.2      & -0.00498               & 0.57028      & 0.13811 \\
0.25     & -0.01112               & 0.69774      & 0.17859 \\
0.3      & -0.02468               & 0.82209      & 0.22846 \\
0.35     & -0.05385               & 0.92445      & 0.28949 \\
0.4      & -0.11798               & 0.98491      & 0.36985 \\
0.42     & -0.16215               & 0.99685      & 0.41056 \\
0.427    & -0.18142               & 0.99984      & 0.42641 \\
\hline
\end{tabular}\end{center}
\caption{Twisted vortex perturbations: parameters of the dispersion relation for $\beta=2.5$}
\label{tab:pertdata25}
\end{table}

\begin{table}[!ht]
\begin{center}
\begin{tabular}{|c||c|c|c|}
\hline
$\omega$ & $\Omega_{\text{min}}^2$ & $\Omega_2^2$ & $k_{\text{min}}$ \\
\hline\hline
0.1      & -0.000155              & 0.284781     & 0.042232 \\
0.5      & -0.007838              & 0.298390     & 0.266817 \\
1        & -0.053988              & 0.364306     & 0.585068 \\
2        & -0.402749              & 0.499033     & 1.356712 \\
3        & -1.195500              & 0.571684     & 2.179592 \\
4        & -2.466004              & 0.618034     & 3.016282 \\
5        & -4.215287              & 0.643379     & 3.855931 \\
\hline
\end{tabular}\end{center}
\caption{Twisted vortex perturbations: parameters of the dispersion relation for $\beta\to\infty$}
\label{tab:pertdataBV}
\end{table}

\begin{table}[!ht]
\begin{center}
\begin{tabular}{|c||c|c|c|c||c|c|c|c|c|}
\hline
$\beta$   & $\omega_{\rm b}$  & $\omega_2$ & $\omega_{\rm b}^{\text{fit}}$ & $\omega_2^{\text{fit}}$ & ${(\Omega^2)}^{(0)}$ & $(\Omega^2)^{(2)}$
& ${(\Omega^2)}^{(0)}_{\text{fit}}$ & $(\Omega^2)^{(2)}_{\text{fit}}$ & $\alpha$\\
\hline\hline
1.25 & 0.13694       & -0.09926  & 0.13694 & -0.09790 & -0.0188 & 0.111 & -0.0188 & 0.0802 & --\\
2.0  & 0.32992       & -0.17824  & 0.32989 & -0.17815 & -0.1088 & 0.488 & -0.1088 & 0.3914 & 0.614 \\
2.5  & 0.42744       & -0.21244  & 0.42744 & -0.21225 & -0.1827 & 0.720 & -0.1826 & 0.6003 & 0.501 \\
\hline
\end{tabular}\end{center}
\caption{Perturbative and numerical data of vortex instability modes. Here, the subscript ``fit'' denotes data obtained by
fitting a parabola to the numerically obtained dispersion relation, close to the minimum.}
\label{tab:pert}
\end{table}

\begin{figure}[!ht]
\noindent\hfil\includegraphics[scale=.3,angle=-90]{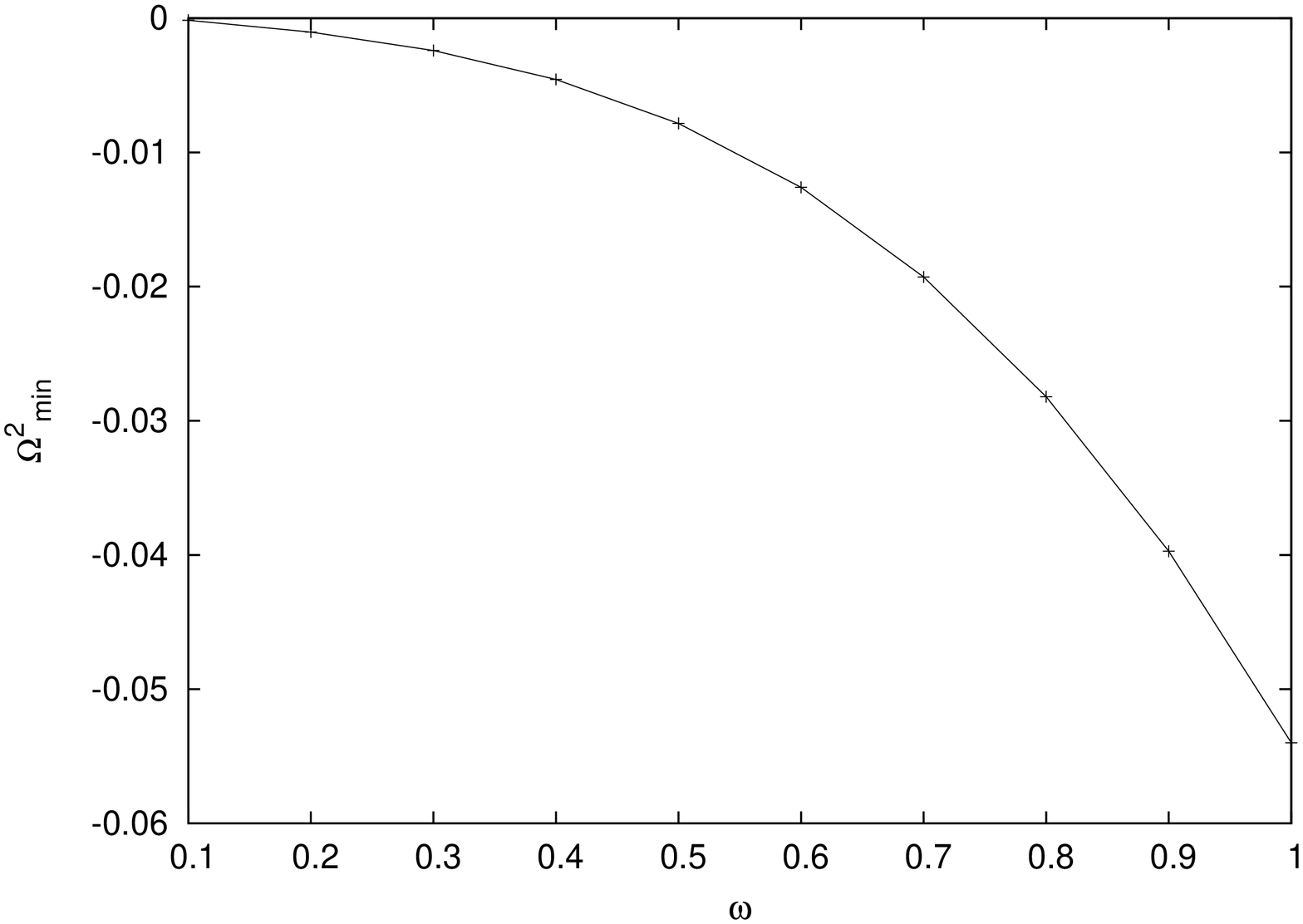} \includegraphics[scale=.3,angle=-90]{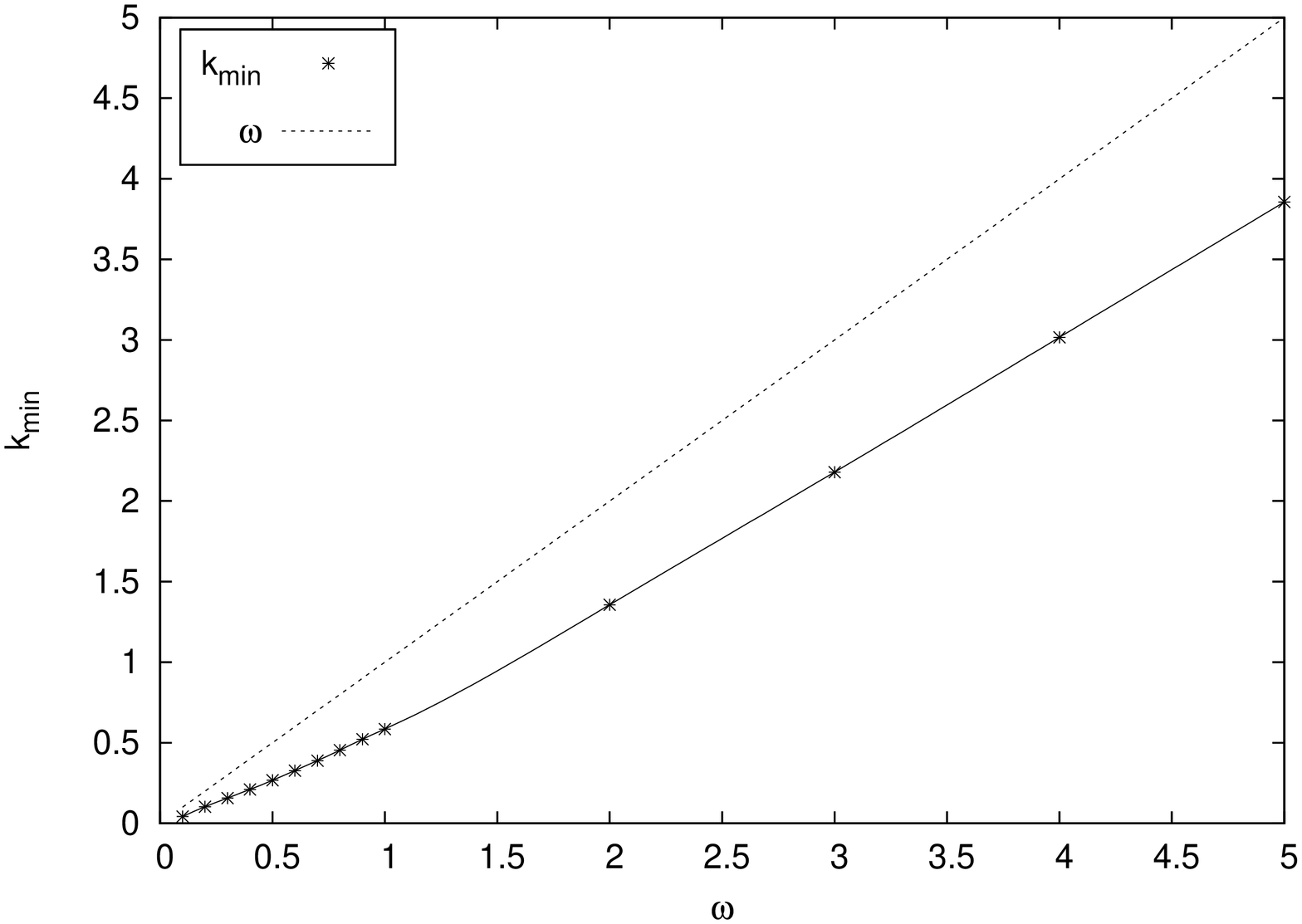}
\caption{Minima of $\Omega^2$ and its location for $\beta\to\infty$ as a function of $\omega$}
\label{fig:omOmBV}
\end{figure}

\begin{figure}[!ht]
\noindent\hfil\includegraphics[scale=.3,angle=-90]{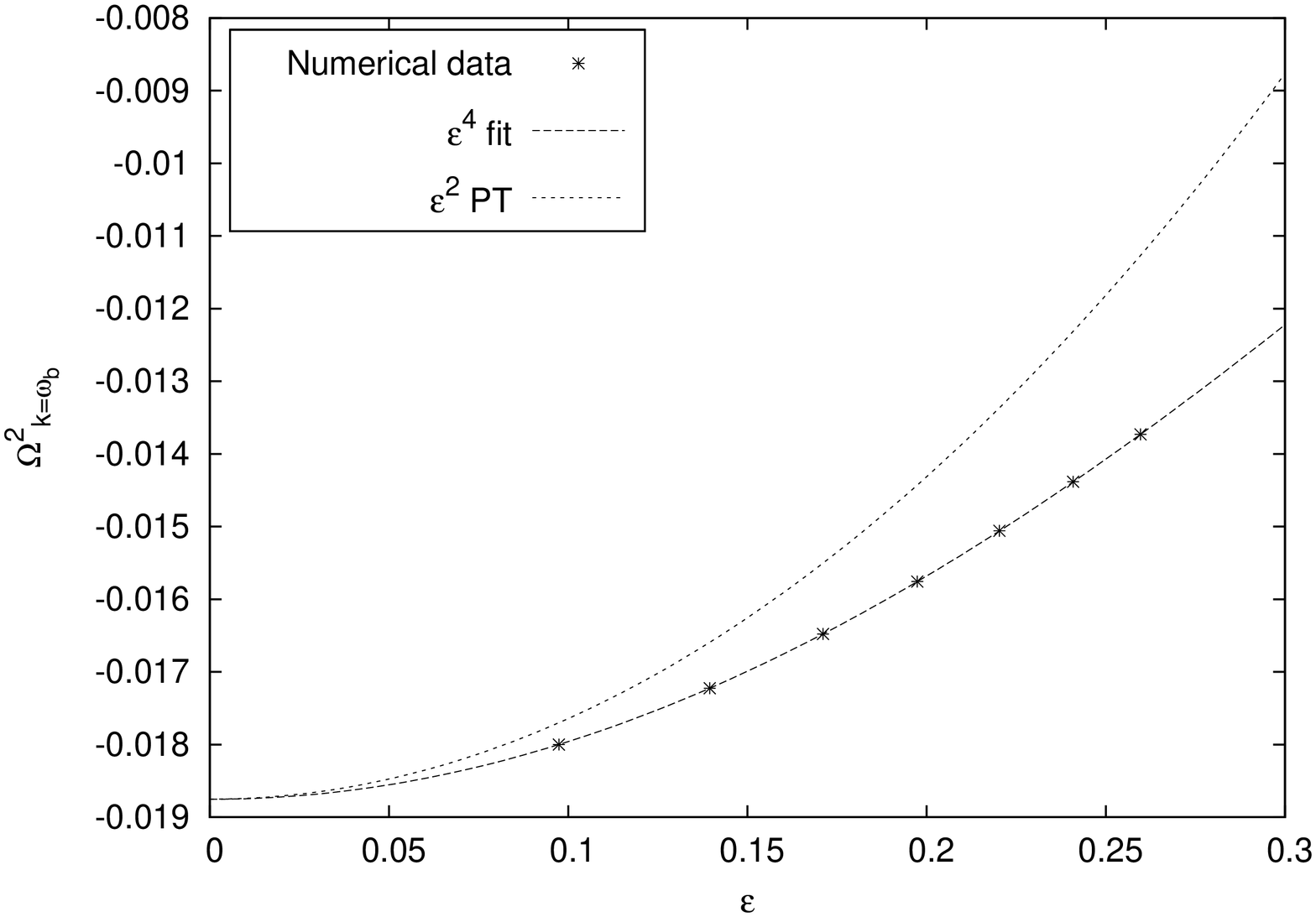} \includegraphics[scale=.3,angle=-90]{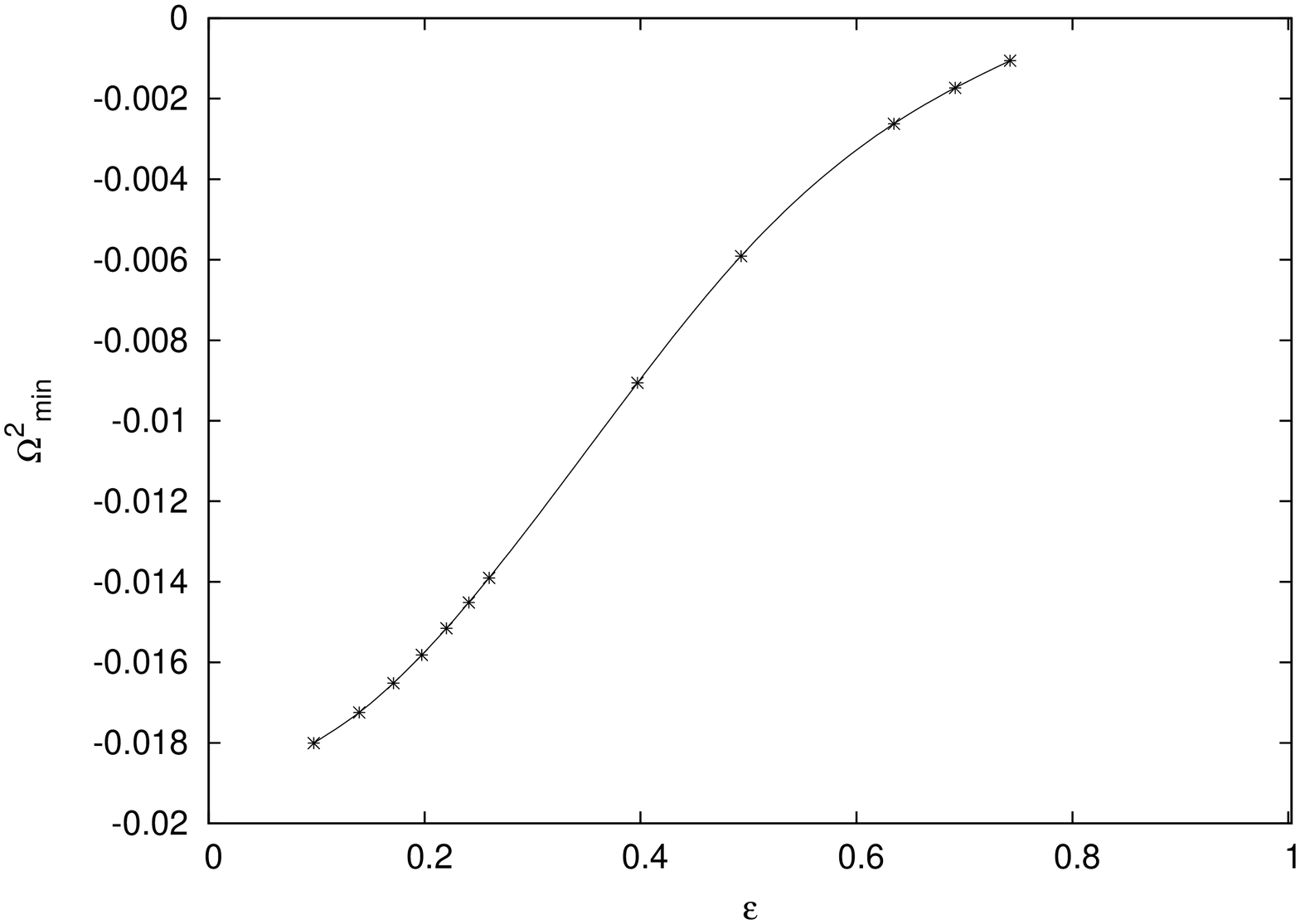}
\caption{$\Omega^2_{k=\omega_{\rm b}}$ and its comparison with PT and the dependence of the minimal $\Omega^2$ on $\epsilon$
for $\beta=1.25$}
\label{fig:pert125}
\end{figure}

It is important to note, that the modes discussed in this section are all part of the physical spectrum. This can be
verified by solving the ghost equation (\ref{eq:FRVghost}). The ghost modes we obtained do not have nodes and correspond
to a positive eigenvalue, therefore there cannot be lower eigenvalue modes. This shows, that the instability
modes presented in this paper are all physical (i.e.\ there are no negative eigenvalue ghost modes which could cancel them).

For the $\beta\to\infty$ case the results of the numerical calculations can be seen (with the parametrization
used in
the $\beta<\infty$ case) in Table \ref{tab:pertdataBV}.
See also the dispersion relation in Figure \ref{fig:disprel}. In this case, there are no ghost mode solutions
(note also that in the regular case, the ghost mode eigenvalues increase for larger values of $\beta$).
The $\beta\to\infty$ data shows that the instability modes persist for all values of $\beta$.
For $\beta\to\infty$ $\omega_{\rm b}\to\infty$, thus the $\beta\to\infty$ data can be
used to study vortices with values of $\omega \ll \omega_{\rm b}$.
Our data indicates that for $\beta$ fixed, the family of unstable modes
parameterized by $k\in[-k_{\rm m},k_{\rm m}]$ also persists for all values of $0<\omega<\omega_{\rm b}$.

\subsection{Bifurcation analysis of the perturbation operator}\label{sec:pertbif}
For values of the twist $\omega\approx\omega_{\rm b}$ we have expanded the vortex backgrounds in a
suitable parameter $\epsilon$. In this paragraph we give the second order expansion of the perturbation operator and its
eigenvalues in the bifurcation parameter $\epsilon$. For details, see Appendix \ref{app:perteqnsSO}.

The leading order equation of the mode pertinent to our problem is
\begin{equation}
  \label{eq:s20}
  D_2^{(0)} s_2 =  (\Omega^2)^{(0)} s_2\,,
\end{equation}
which yields the unstable mode of the embedded ANO vortex. Let us note here, that for $k=\omega$,
this instability mode is $z$-independent. For other values of $k$, its $z$-dependence is
harmonic.

Let us now look at the first order correction to the eigenvalue.
This correction, as known from quantum mechanics is given as
\begin{equation}
  \label{eq:DOmeg1}
  (\Omega^2)^{(1)} = \frac{1}{\|s_2\|^2} \left( s_2, D_2^{(1)} s_2 \right) = 0.
\end{equation}
Therefore, to obtain an $\omega$-dependent eigenvalue, one must go to the second order in perturbation theory.
Contributions quadratic in $\epsilon$ come from two sources: from ${\cal O}(\epsilon^2)$ terms in the perturbation
operator (i.e.\ corrections to the background functions of order $\epsilon^2$), and from second order perturbation theory,
\begin{equation}
  \label{eq:SOpert}
  (\Omega^2)^{(2)} = (\Omega^2)^{(2)}_B + (\Omega^2)^{(2)}_P.
\end{equation}
The first term is simply
\begin{equation}
  \label{eq:Om2B}
  (\Omega^2)^{(2)}_B = \frac{1}{\|s_2\|^2} \left( s_2, D_2^{(2)} s_2 \right) = M_{22} + M_{22}^k(k-\omega_{\rm b})
\end{equation}
with
\begin{equation}
  M_{22} = \frac{1}{\|s_2\|^2} \left( s_2, \left[ -\frac{2(\ell-na^{(0)})na^{(2)}}{r^2}+2\beta f_1^{(0)}f_1^{(2)}+(2\beta+1)(f_2^{(1)})^2\right] s_2 \right)
\end{equation}
and
\begin{equation}
  M_{22}^k = \frac{1}{\|s_2\|^2}\left( s_2, \left[ -2\omega_2 + 2 \omega_{\rm b} a_3^{(2)}\right] s_2\right).
\end{equation}

To calculate the other contributions, let us first introduce the following notations ($\delta A_0$ component
dropped -- it is decoupled in all orders; see also Appendix \ref{sec:local}.)
\begin{equation}
  \label{eq:P2modes}
\begin{aligned}
  \psi_2\,  &:= (0,0,s_2,0,0,0,0)^T \,,\\
  \psi_2^\dagger  &:= (0,0,0,s_2^*,0,0,0)^T \,,\\
  \psi_S &:= (s_0^S, s_0^S, 0,0,a_0^S,a_0^S,0)^T \,,\\
  \psi_A &:= (s_0^A, -s_0^A, 0,0,a_0^A,-a_0^A,0)^T \,,\\
  a_3    &:= (0,0,0,0,0,0,a_3)^T \,.
\end{aligned}
\end{equation}
In Eq.\ (\ref{eq:P2modes}), $\psi_2$ is the unstable mode of the embedded ANO vortex (see Eq.\ (\ref{eq:s20})) and
$\psi_2^\dagger$ is its conjugate mode (it can be seen easily, that the equation of $s_2^*$
is the same as that of $s_2$ with the transformation $k\to-k$). $\psi_S$ and $\psi_A$ are the symmetric and antisymmetric bound modes
of the ANO vortex, respectively, while $a_3$ is another bound mode of the ANO vortex, satisfying the same equation as ghosts.

The dispersion relation of these modes (in the same order as in Eq.\ (\ref{eq:P2modes})) is given as
\begin{equation}\label{eq:P2mEig}
\begin{aligned}
    \Omega^2_{s2}\, &= \lambda_2^2 + (k-\omega)^2 \,,\\
    \Omega^2_{s2*}  &= \lambda_2^2 + (k+\omega)^2 \,,\\
    \Omega^2_{S}\,  &= \lambda_S^2 + k^2 \,,\\
    \Omega^2_{A}\,  &= \lambda_A^2 + k^2 \,,\\
    \Omega^2_{a3}\, &= \lambda_{a3}^2 + k^2 \,,\\
\end{aligned}
\end{equation}
with $\lambda_i$ being constants (lowest eigenvalues).
For second order perturbation theory, the following matrix elements are needed,
\begin{equation}
\label{eq:P2mat}
\begin{aligned}
  M_{22*} &= \frac{1}{\|s_2\|^2}\left(s_2, U_2^{(1)} s_2^*\right) = 0 \,,\\
  M_{2S}  &= \frac{1}{\|s_2\| \|\psi_S\|}\left[
                 \left(s_2, (V^{(1)}+{V'}^{(1)})s_0^S\right) + \left(s_2, (A_2^{(1)}+{A_2'}^{(1)})a_0^S\right) \right]\,,\\
  M_{2A}  &= \frac{1}{\|s_2\| \|\psi_A\|}\left[
                 \left(s_2, (V^{(1)}-{V'}^{(1)})s_0^A\right) + \left(s_2, (A_2^{(1)}-{A_2'}^{(1)})a_0^A\right) \right]\,,\\
  M_{23}  &= \frac{1}{\|s_2\| \|a_3\|} \left( s_2, B_2^{(1)} a_3\right)\,.
\end{aligned}
\end{equation}
The part of the ${\cal O}(\epsilon^2)$ correction to the eigenvalue due to perturbation theory can be written as
\begin{equation}
\label{eq:Om2P}
  (\Omega^2)^{(2)}_P = \frac{|M_{2S}|^2}{\lambda_S^2-2\omega_{\rm b}^2+2k\omega_{\rm b}} + \frac{|M_{2A}|^2}{\lambda_A^2-2\omega_{\rm b}^2+2k\omega_{\rm b}}
                      + \frac{|M_{23}|^2}{\lambda_{a3}^2-2\omega_{\rm b}^2+2k\omega_{\rm b}}
\end{equation}
The contribution to the perturbation functions will be evaluated similarly using the formulas
well known from elementary quantum mechanics in Subsection \ref{sec:perteig}.

Looking at the potentials in the equations of $\psi_S$ and $\psi_A$ (see Figure \ref{fig:ANO-pot} in Appendix \ref{sec:local}),
one can easily understand why the mode
$\psi_S$ becomes quasi-bound
(instead of bound) for values of $\beta$ above $1.5$ (see Appendix \ref{sec:local}). The contribution of these modes to PT has to be taken into
account when calculating the perturbation
(\ref{eq:Om2P}) of the eigenvalue. This can be done as follows: the $s$-component of the quasi-bound mode
varies slowly with the energy in a
given interval, while the $a$-component varies rapidly. This makes it possible, to smoothen the contribution
by using a low-lying quasi-bound mode,
with the $a$-component taken to be zero
and reading off the integral of the state density over the energy interval, in which the $s$-component is
nearly constant, from the perturbation
functions, and using this correction factor $\alpha$ to calculate the energy correction:
\begin{equation}
\label{eq:Om2Pqb}
  (\Omega^2)^{(2)}_P = \frac{\alpha|M_{2S}|^2}{\lambda_S^2-2\omega_{\rm b}^2+2k\omega_{\rm b}} + \frac{|M_{2A}|^2}
  {\lambda_A^2-2\omega_{\rm b}^2+2k\omega_{\rm b}}
                      + \frac{|M_{23}|^2}{\lambda_{a3}^2-2\omega_{\rm b}^2+2k\omega_{\rm b}}
\end{equation}
with the matrix element$M_{2S}$ calculated from the smoothed $\psi_S$, and $\alpha$ is obtained from a comparison of the
perturbative and exact eigenmodes.

The eigenvalue $\Omega^2$ at $k=\omega_{\rm b}$ as a function of $\epsilon$ (near $\omega=\omega_{\rm b}$) can easily
be calculated with perturbation theory as
outlined in the previous section. Second order perturbation theory predicts
\begin{equation}
  \label{eq:pertE}
  \Omega^2_{k=\omega_{\rm b}} = (\Omega^2)^{(0)} + \epsilon^2 (\Omega^2)^{(2)}.
\end{equation}
Values of $(\Omega^2)^{(0)}$ and $(\Omega^2)^{(2)}$ obtained with the perturbation theory methods of the
previous section and exact (numerical)
values are compared in Table \ref{tab:pert} and Figure \ref{fig:pert125}.
Similarly to the case of the vortex backgrounds close to the bifurcation, there is a good agreement
between the exact eigenvalues and the perturbative ones, for moderate values of $\epsilon$, i.e.\ approximately for
$\epsilon \le 0.2$. This is an indication that the instability mode of the twisted vortex can be regarded as a
deformation of the instability mode of the embedded ANO vortex.

\subsection{Perturbative calculation of the eigenfunction corresponding to the instability}\label{sec:perteig}
In this section, we present the perturbation functions $s_{1,\ell},s_{1,-\ell}^*,s_{2,\ell},s_{2,-\ell}^*,a_\ell,a_{-\ell}^*,a_{3,\ell}$.
We will also compare these with their first order perturbative expressions near the bifurcation point.

Let us first note, that for vortices close to the bifurcation (i.e. $\omega$ close to $\omega_{\rm b}$, $\epsilon \ll 1$), the
lowest lying negative eigenvalue is close to $k=\omega$. Let us now choose a vortex which can still be treated perturbatively,
but it is not very close to an embedded ANO vortex, eg.\ $\omega=0.13$ for $\beta=1.25$, and examine its instability
mode for $k=\omega$.

\begin{figure}[!ht]
\noindent\hfil\includegraphics[scale=.5,angle=-90]{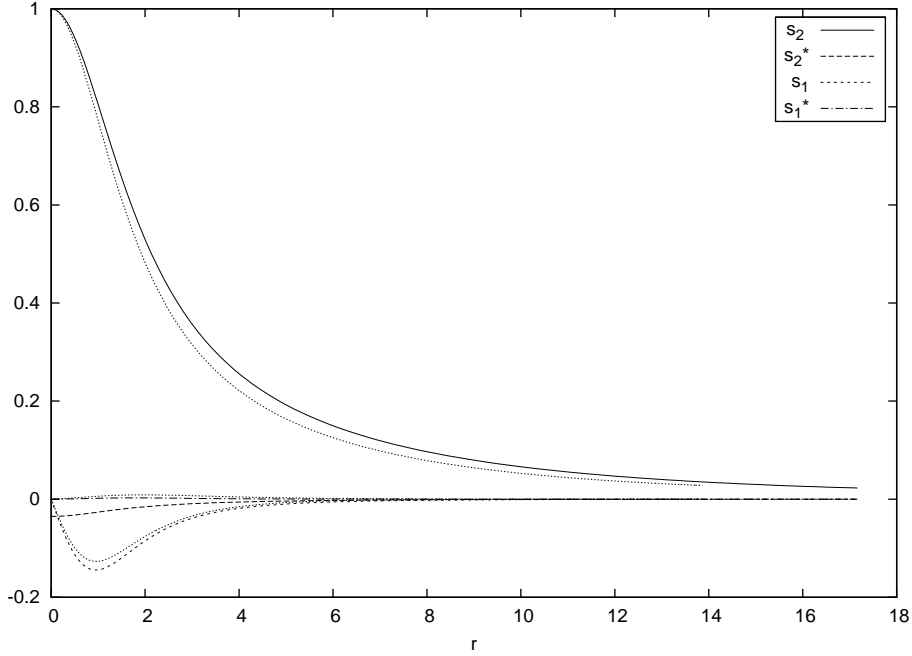}
\caption{The twisted vortex instability mode, scalar field components for $\beta=1.25$, $\omega=0.13$. Perturbative results
for $s_1$, $s_1^*$ and $s_2$  are shown with dotted lines, $s_2^*\equiv 0$  to this order.}
\label{fig:pertF}
\end{figure}

\begin{figure}[!ht]
\noindent\hfil\includegraphics[scale=.5,angle=-90]{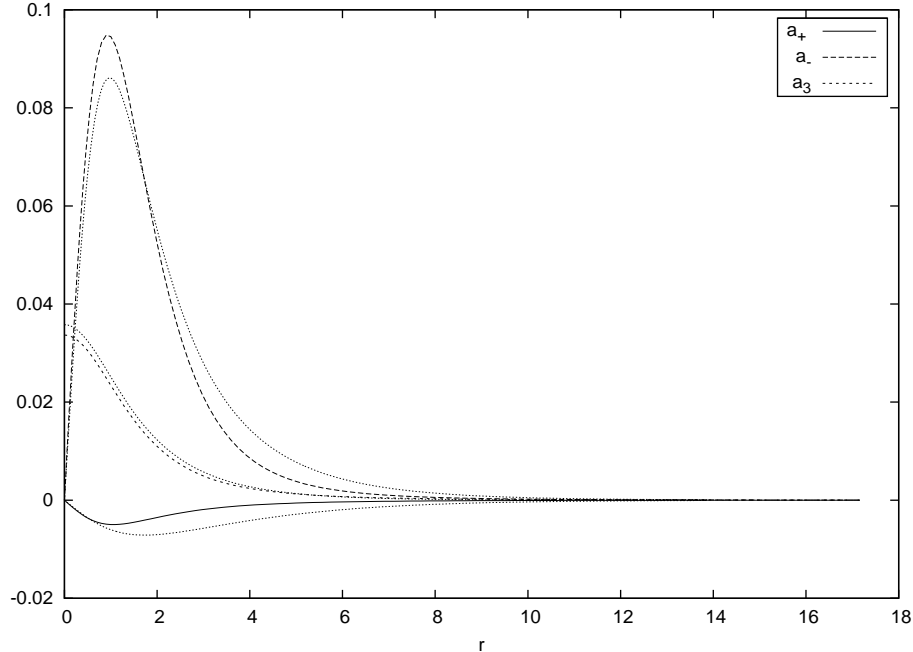}
\caption{The twisted vortex  instability mode, vector field components for $\beta=1.25$, $\omega=0.13$. Perturbative results
  are shown with dotted lines.}
\label{fig:pertA}
\end{figure}

In the previous subsection, we have applied perturbation theory to obtain the eigenvalues of the perturbation
operator. Let us now apply PT to evaluate the wave
function, i.e.\ the unstable mode of the twisted vortex. The unperturbed wave function is simply the instability
mode of the embedded ANO vortex,
\begin{equation}\label{eq:Ppsi0}
  \psi^{(0)} = \psi_2 = (0,0,s_2,0,0,0,0)^T \,.
\end{equation}
Let us now write up the first order perturbations to this mode,
\begin{equation}\label{eq:Ppsi1}
  \psi^{(1)} = -N\frac{M_{2S}}{\lambda_S^2-2\omega_{\rm b}^2+2k\omega_{\rm b}}\psi_S
  -N\frac{M_{2A}}{\lambda_A^2-2\omega_{\rm b}^2+2k\omega_{\rm b}}\psi_A
  -N\frac{M_{23}}{\lambda_3^2-2\omega_{\rm b}^2+2k\omega_{\rm b}}a_3
\end{equation}
where $N={\|\psi_2\|}$. The first order wavefunction and the exact instability mode exhibit a good agreement, even
for not so small values of $\epsilon$ (see Figures \ref{fig:pertF} and \ref{fig:pertA}; in that case $\epsilon=0.264$).


Note that the perturbations depicted in Figures \ref{fig:pertF}--\ref{fig:pertA}, correspond to a
quite large value of $\epsilon$, and even so the instability
mode components are still strongly dominated by $s_2$. This indicates that
the instability of the twisted vortex proceeds similarly to that of the embedded ANO one,\
i.e.\ the vortex starts to expand nearly homogeneously.
To support this conclusion further, we calculate the perturbations of the current of the vortex.
The components, which receive perturbations, are
\begin{equation}
  \label{eq:Pcurr0a}
  \delta j^3_0 = e^{i\ell\vartheta}e^{i(\Omega t-kz)} \Omega \left\{ f_1 (s_{1,-\ell}^* - s_{1,\ell}) - f_2
    (s_{2,-\ell}^* - s_{2,\ell}) \right\} \,,
\end{equation}
where, $s_{1,-\ell}^* = {\mathcal O}(\epsilon)$, $s_{1,\ell} = {\mathcal O}(\epsilon)$, and $f_2 = {\mathcal
  O}(\epsilon)$, i.e.\ this perturbation does not change the zero component of the current in the leading order, and
\begin{equation}
  \label{eq:Pcurr3a}
  \begin{aligned}
  \delta j_3^3 = e^{i\ell\vartheta}e^{i(\Omega t-kz)}
    &\left\{ ( k+2\omega a_3) f_1 s_{1,\ell} - (k+2\omega(a_3-1))f_2 s_{2,\ell} \right. \\
    +&\left.( -k+2\omega a_3) f_1 s_{1,-\ell}^* - (-k+2\omega(a_3-1))f_2 s_{2,-\ell}^* + 2(f_1^2 - f_2^2) a_{3,\ell}
    \right\} \,,
  \end{aligned}
\end{equation}
where $s_{1,-\ell}^* = {\mathcal O}(\epsilon)$, $s_{1,\ell} = {\mathcal O}(\epsilon)$, $a_{3,\ell} = {\mathcal
  O}(\epsilon)$ and $f_2 = {\mathcal O}(\epsilon)$ therefore $\delta j_3^3 = {\mathcal O}(\epsilon)$ too.

The deviation of both the eigenvalues and the eigenfunctions of twisted strings is of $\mathcal{O}(\epsilon^2)$
with respect to those of the ANO ones. This suggests that the eigenmode of the ANO string is an energy lowering
perturbation of the twisted one.
The energy, $\Omega^2 = \left( \psi_2 , D_2 \psi_2 \right)$, of the embedded ANO instability mode, $s_2$,
on a twisted vortex background can be expanded as
\begin{equation}
  \Omega^2 = (\Omega^2)^{(0)} + \epsilon^2 (\Omega^2)^{(0)}\,.
\end{equation}
Using the expansion of the perturbation operator
\begin{equation}
  D_2 = D_2^{(0)} + \epsilon^2 D_2^{(2)} \,,
\end{equation}
yields
\begin{equation}
  (\Omega^2)^{(2)} = M_{22} + M_{22}^k (k-\omega_{\rm b})\,.
\end{equation}
Table \ref{tab:omega2p} shows clearly the important fact
that for not too large values of $\epsilon$, the perturbation $s_2$ still lowers the energy.
The $\epsilon_{\text{max}}$ values were calculated with $k=\omega_{\rm b}$.

\begin{table}
\begin{center}
\begin{tabular}{|c||c|c|c|c|}
\hline
$\beta$ & $M_{22}$  & $M_{22}^k$  & $(\Omega^2)^{(0)}$ & $\epsilon_{\text{max}}$ \\
\hline \hline
$1.25$  & $0.6883$  & $0.2934$   & $-0.0188$          & $0.1653$ \\
$2$     & $1.2442$  & $0.5990$   & $-0.1088$          & $0.2957$ \\
$2.5$   & $1.4843$  & $0.7270$   & $-0.1827$          & $0.3396$ \\
\hline
\end{tabular}
\end{center}
\caption{Data for the calculation of the energy of the mode $\psi_2$ on a twisted vortex background}
\label{tab:omega2p}
\end{table}

\subsection{Properties of the instability modes}\label{sec:prop}

Let us recapitulate first, that, as shown in Subsection \ref{sec:linpert}, the obtained unstable eigenmodes are all physical.
An examination of the above calculated eigenvalues and the corresponding eigenfunctions shows clearly that the obtained
instability eigenmode is a deformation of the instability eigenmode of the embedded ANO vortices already mentioned in
Section \ref{sec:BGbif} (See also Appendix \ref{sec:local}), in the sense that a for a given value of the $z$-direction
wavenumber $k$, the corresponding eigenmode $\Psi$ and eigenvalue $\Omega$ are smooth functions of the twist $\omega$, which, for
$\omega=\omega_{\rm b}$ reproduce the eigenmodes and eigenvalues of the embedded ANO vortices.
In Subsection \ref{sec:perteig} we presented a description of this
deformation based on perturbation theory, with the unperturbed solutions being the eigenmodes of the embedded ANO vortex.

The most important result is that for each $k\in [-k_{\rm m},k_{\rm m}]$ there is one eigenmode corresponding to a negative eigenvalue
(here $k_{\rm m}$, of course, depends on $\omega$).
In this section we would like to summarize the properties of these eigenmodes, and speculate about the physical consequences of them.

Firstly, let us briefly analyze the properties of the instability modes. Let us point out that even for values of
$\omega$
quite far from the bifurcation $\omega_{\rm b}$
($\epsilon\approx 0.2\dots 0.3$), the unstable mode is dominated by the $s_2$ component (see also Section \ref{sec:perteig} and
Figures \ref{fig:pertF}--\ref{fig:pertA} therein), and the lowest lying eigenvalue is close to $k=\omega$. If
$k=\omega$ then the dominant $s_2$ component of the unstable mode is $z$-independent. In the previous subsection we have shown,
that for $\omega\approx\omega_{\rm b}$ (see Table \ref{tab:omega2p} for critical values of $\epsilon$ at which the energy of this
perturbation becomes positive) the unstable mode of the ANO vortex is also an energy lowering perturbation for the twisted one,
albeit not an eigenmode. In this way we can explicitly construct a $z$-independent energy-lowering perturbation of the twisted string,
which is an indication that homogeneous expansion of twisted strings close to the bifurcation is possible.

The deviation from $k_{\text{min}}=\omega$ occurs at small values of $\omega$, far from the bifurcation,
$\omega=\omega_{\rm b}$, i.e.\ the $z$-dependence of the instability eigenmode becomes important when the vortices have
already quite expanded in the lower scalar field component. In our opinion, this is another indication that the
instability corresponds to an expansion of the string.

In order to see how the string can expand in case a completely homogeneous expansion is not possible, let us take a perturbation
of the form of a wave packet, centered
around $z=0$ in coordinate space and around $k=k_{\text{min}}$ in momentum space. Then, looking at the Fourier transform of such a wave packet one gets
$
\Delta x \Delta k = 1
$
for the width of the wave packet in coordinate and momentum space, respectively. Let the amplitude of the packet be $A$. Then, the energy of this packet
can be calculated as
\[
  E = |A|^2 \left[ \Omega_{\text{min}}^2  + \Omega_2^2 (\Delta k)^2 \right] + \dots \,,
\]
which can be negative if $\Delta k$ is sufficiently small. However, for the wave packet to make the current grow (locally make the vortex
resemble a vortex of lower $\omega$, i.e.\ let the vortex locally get diluted), one needs
\[
\Delta x < \lambda = \frac{2\pi}{k_{\text{min}}}\,,
\]
which gives \[ \Delta k \approx \frac{k_{\text{min}}}{2\pi} \,.\]
Substituting this into the expression for the energy, at least close to the bifurcation point, one can see that
\[
E \approx |A|^2\left(\frac{\Omega_2^2}{4\pi^2}-1\right)k_{\text{min}}^2 < 0 \,.
\]
We have also checked, using the values of $\Omega_{\text{min}}^2$
and $\Omega_2^2$ obtained above, that for $\beta=2$ for all values of $\omega$ available for our numerical methods the energy contribution of
the lump can be negative,
which suggests that the instability of the vortex corresponds to an expansion instability, similarly to the case of the embedded ANO vortices.
This result makes it plausible that a local (lump-like) spreading of the vortex resulting in a configuration with diluted magnetic flux is possible.
Forthermore, a (local) expansion of the vortex can form out of a general initial perturbation of the string, because the eigenmodes which are almost
$z$-independent are the ones which are close to mode corresponding to the largest negative eigenvalue, and therefore they are the ones that grow most
rapidly in time, and a general initial perturbation (like the lump discussed above) overlaps with these modes.

The existence of the $z$-independent energy-lowering perturbations for strings close to the bifurcation and local lumps for
all strings independently of $\omega$ clearly show that at least at the outset of the instability an expansion of the string is
possible. In the case of embedded ANO vortices it is known \cite{hin,akpv}, that they
are unstable against expansion.
The similarity of the eigenmodes of twisted and embedded ANO strings supports that at least initially, twisted
strings also start to expand.
However, in the linear approximation, the perturbations of the current components, Eq.\ (\ref{eq:Pcurr0a}) and 
(\ref{eq:Pcurr3a}), are harmonic in $z$, and thus the conserved current of the twisted
string remains localized. This changes the dynamics of the expansion, compared to that of
the embedded ANO strings considerably;
the expansion cannot go on indefinitely until the current is carried away by radiation. Radiation effects only occur
in higher orders of perturbation theory, and thus
the framework of the present paper, i.e.\ linearization, is not sufficient to draw
definite conclusions about the long term dynamics of the instability, and the resulting final state.

The stability of the global version of current carrying strings, in a theory with two scalar fields, has been analyzed
in Ref.\ \cite{Lemperiere}. It has been found that the string is stable against homogeneous and only $z$-dependent
perturbations, and above a critical value of the (global) current unstable against time dependent ones.  For twisted
vortices in the semilocal model, it is the magnetic flux in the string that is repsonsible for the spreading
instability.
%
%

Ref.\ \cite{volkov} suggests another scenario for the instability of semilocal twisted strings, based on the harmonic
$z$-dependence of the $SU(2)$-current flowing in the string (see Eqns.\ (\ref{eq:Pcurr0a}) and (\ref{eq:Pcurr3a})
in the previous subsection), and the fact that a larger current corresponds to a more expanded (lower $\omega$) string.
There, an analogy with the Plateau-Rayleigh instability of a fluid stream (droplet formation) is emphasized.
The analogue of the Plateau-Rayleigh instability, as suggested by Ref.\ \cite{volkov} is the break-up of the string
into small, droplet-like string segments. 

As mentioned above, there are good reasons to believe thet for not too large values of the current, twisted strings start to
expand initially. For large values of the current where the nearly homogeneous expansion is not an energy lowering perturbation,
the width of the string is already quite large, therefore, in neither case can one conclude that the string will initially
start to break up.

Recently, in Ref.\ \cite{vorton} the dynamics of twisted vortex loops (vortons) has been studied in a model with two scalar fields.
There, it has been found that such a loop is stable against axially symmetric perturbations (eg.\ expansion), and unstable against
non-axially symmetric ones. Longer time evolution shows that the loop breaks up into smaller pieces. In our opinion however,
this result in itself is not decisive to support the breakup scenario of twisted strings in the semilocal model.
Besides the different initial dynamics, anoter reason why breakup of the string is unlikely is that, in order for the string to
break up, string ends and droplet-like configurations must form. 
Clearly if a string breaks, its magnetic flux must end in something like a magnetic monopole resulting in a large energy contribution.
We expect the situation to be rather different when non-Abelian gauge fields are present,
since then a possible configuration for a string piece would be the famous dumbbell of Nambu \cite{nambu}.

Another interesting question examined in Ref. \cite{volkov} is the stability of solutions with periodic boundary conditions at $z=0,L$
(this is an approximation of a vortex loop, if $L$ is large).
In the semilocal model, periodicity constrains the twist, so that $\omega L$ has to be an integer multiple
of $2\pi$,
\begin{equation}\label{eq:perconstra}
  \omega L = 2\pi p\,,\qquad p\in \mathbb{Z}\,.
\end{equation}
In this case, for not very small values
of $\omega$ (i.e.\ not very far from the bifurcation), the $k=\omega$ mode corresponds to a negative eigenvalue, therefore these
short strings with periodic boundary conditions are unstable in the semilocal model.

An interesting observation of Ref.\ \cite{volkov} is, that considering twisted strings as solutions
of the $SU(2)\times U(1)$ (electroweak) gauge theory, the instability found in the semilocal model can be excluded
by imposing periodic boundary conditions. In the electroweak model,
with a suitable gauge transformation involving the electromagnetic and the third $SU(2)$ component vector fields, the $z$-dependence
of the string can be gauged away and therefore periodic solutions of arbitrary length exist. However
such periodic strings are {\sl not gauge equivalent} anymore to a piece of a twisted string, the
latter being
a discontinuous configuration with infinite energy. Looking at the gauge transformation connecting the twisted and the untwisted
string, the generator of this transformation is proportional to $z$. When $z$ is made periodic, it ceases to be a univalued function,
and thus, the vector field component proportional to $\d z$ ceases to be a pure gauge.

The gauge transformation which removes the twist of the vortex in the gauged $SU(2)\times U(1)$ model renders the homogeneously
expanding mode $z$-dependent, and thus, this perturbed solution is excluded by the periodicity condition. This means that these
small periodic strings
are stable with respect to purely $U(1)$ perturbations, however, perturbations in the $SU(2)$ non-Abelian sector can still cause
instabilities.
Whether such instabilities exist, could only be decided considering all perturbations in the $SU(2)\times U(1)$ gauge
theory, as in the case of electroweak strings \cite{GHew}. Ref.\ \cite{GHew} performs the linearized stability analysis of electroweak
Z-strings. There, Z-stings are found to be unstable unless $\beta$ is small or $\theta_{\rm W}$ is large, and the instability mode is
$z$-independent, occurs in the non-Abelian sector, and is decoupled from the other field components. However, these properties of the
perturbation equations cannot be generalized to the case of the two-component strings of Ref.\ \cite{volkov}.

\section{Conclusion}
We have found that twisted strings in the $SU(2)_{\text{global}}\times U(1)$ extended
Abelian Higgs (semilocal) model possess a family of unstable modes, parametrized by their
wave number in the $z$ direction, $k\in[-k_{\rm m},k_{\rm m}]$.
This result is in agreement with that of Ref.\ \cite{volkov}.
We have also obtained a semi-analytical description of twisted strings based on a suitable perturbative
expansion near their bifurcation point with the embedded
ANO solutions.
The perturbative solution provides a surprisingly good quantitative approximation for twisted
strings themselves, as well as for the instability eigenvalues and eigenfunctions.
This way the instability modes of twisted strings are obtained as deformations of those of the
embedded ANO solutions. 

Based on the similarity of the unstable modes of twisted and embedded ANO strings, we suggest that the 
initial time dynamics is also similar, i.e.\ at the outset of the instability twisted strings start to expand
(nearly) homogeneously, just like embedded ANO ones.
This idea is also supported by the fact, that the (homogeneous) lowest energy eigenmode
of the embedded ANO string is also an energy lowering perturbation for twisted strings with
$\omega$ not far from the bifurcation $\omega_{\rm b}$, although not an eigenmode.
For values of the twist $\omega \ll \omega_{\rm b}$, the formation of a growing
lump leading to a local expansion of the string,
is also an energy lowering perturbation. This suggests that for these strings
(where the condensate in the core is already quite sizable),
a local expansion is thus a plausible scenario for the initial dynamics of the instability.
Twisted strings carry a global current. Although the initially homogeneous distribution of this current is changed
into a $z$-dependent one, the current remains localized in leading order perturbation theory.
Therefore the expansion of the string cannot proceed indefinitely, unless higher order radiation effects do
indeed remove the current from the string.

\section*{Acknowledgments}
This work has been supported by the OTKA Grant Nos. K61639 and NI68228.
The authors would like to thank Zal\'an~Horv\'ath and Philippe Grandcl\'ement
for discussions.

\appendix

\section{Details of the perturbation equations}\label{app:perteqns}
In this section we present some of the omitted details of the calculations. Although in other parts of the paper we
have only studied the $n=1$, $m=0$ case, here we present the general, arbitrary $n$ and $m$ case.
\subsection{The perturbation operator of the twisted vortices}
Let us apply the Fourier transform (\ref{eq:Fourier}) to the equations (\ref{eq:FRVlag2b}). A purely harmonic mode can be written
in detail as
\begin{equation}
  \label{eq:FRV-fouriermodes}
  \begin{aligned}
    \delta \phi_1 (z,t;x_i) &= e^{i(\Omega t-k z)} \delta \phi_1 (k,\Omega;x_i) \\
    \delta \phi_2 (z,t;x_i) &= e^{i(\Omega t-(k-\omega) z)} \delta \phi_2 (k,\Omega;x_i) \\
    \delta A_\mu (z,t;x_i) &= e^{i(\Omega t-k z)} \delta A_\mu (k,\Omega;x_i)
  \end{aligned}
\qquad
  \begin{aligned}
    \delta \phi_1^* (z,t;x_i) &= e^{i(\Omega t-k z)} \delta \phi_1^* (-k,-\Omega;x_i) \\
    \delta \phi_2^* (z,t;x_i) &= e^{i(\Omega t-(k+\omega) z)} \delta \phi_2^* (-k,-\Omega;x_i) \\
    {}\\
  \end{aligned}
\end{equation}
with the index $i$ running over 1,2.
The variables $A_\mu$ are real functions, therefore
\begin{equation}
  A_\mu(k,\Omega,x_i) = A_\mu^*(-k,-\Omega,x_i).
\end{equation}
Substituting these into equations (\ref{eq:FRVlag2b}) yields the perturbation operator $\mathcal{M}$ of equation (\ref{eq:FRV-fourier}):
\begin{equation}
  \label{eq:fourM}
  \mathcal{M}=\begin{pmatrix}
    \mathcal{D}_1      & \mathcal{U}_1   & \mathcal{V}_1       & \mathcal{V}_1'   & \mathcal{A}_1   & \mathcal{B}_1   & 0 \\
    \mathcal{U}_1^*    & \mathcal{D}_1^* & \mathcal{V}_1'{}^{*} & \mathcal{V}_1^*  & \mathcal{A}_1^* & \mathcal{B}_1^* & 0 \\
    \mathcal{V}_2      & \mathcal{V}_2'  & \mathcal{D}_2       & \mathcal{U}_2    & \mathcal{A}_2   & \mathcal{B}_2   & 0 \\
    \mathcal{V}_2'{}^* & \mathcal{V}_2^* & \mathcal{U}_2^*     & \mathcal{D}_2^*   & \mathcal{A}_2^* & \mathcal{B}_2^* & 0 \\
    \mathcal{A}_1^*    & \mathcal{A}_1   & \mathcal{A}_2^*     & \mathcal{A}_2     & \mathcal{D}_3   & 0              & 0 \\
    \mathcal{B}_1^*    & \mathcal{B}_1   & \mathcal{B}_2^*     & \mathcal{B}_2     & 0              & \mathcal{D}_3   & 0 \\
    0                  & 0               & 0                  & 0                 & 0               & 0 & \mathcal{D}_3 \\
  \end{pmatrix}
\end{equation}
with
\[\begin{aligned}
\mathcal{D}_1    &= k^2 +\mathcal{D}_s + 2 i A_i \partial_i + 2 k A_3 +\mathcal{W}_1\,,\\
\mathcal{D}_1^*  &= k^2 +\mathcal{D}_s - 2 i A_i \partial_i - 2 k A_3 +\mathcal{W}_1\,,\\
\mathcal{D}_2    &= (k-\omega)^2 +\mathcal{D_s} + 2i A_i \partial_i - 2(\omega-k)A_3 +\mathcal{W}_2\,, \\
\mathcal{D}_2^*  &= (k+\omega)^2 +\mathcal{D_s} - 2i A_i \partial_i - 2(\omega+k)A_3 +\mathcal{W}_2\,, \\
\mathcal{D}_3   &= k^2 - \partial_i^2 + 2 |\phi|^2\,,
\end{aligned}\]
where
\[
\mathcal{D}_s = - \partial_i^2 + A_i^2 + A_3^2 + \beta (|\phi|^2-1) \qquad \mathcal{W}_i = (\beta+1)\phi_i\phi_i^*
\]
and
\[\begin{aligned}
\mathcal{U}_1  &= (\beta-1)\phi_1^2 \\
\mathcal{V}_1  &= (\beta+1)\phi_1\phi_2^* \\
\mathcal{V}_1' &= (\beta-1)\phi_1\phi_2 \\
\mathcal{A}_1  &= 2 A_i \phi_1 + 2 i\partial_i\phi_1 \\
\mathcal{B}_1  &= 2 A_3 \phi_1
\end{aligned}\qquad
\begin{aligned}
\mathcal{U}_2  &= (\beta-1)\phi_2^2 \\
\mathcal{V}_2  &= (\beta+1)\phi_2\phi_1^* \\
\mathcal{V}_2' &= (\beta-1)\phi_2\phi_1\\
\mathcal{A}_2  &= 2 A_i \phi_2 + 2 i\partial_i\phi_2 \\
\mathcal{B}_2  &= 2 A_3 \phi_2
\end{aligned}\]
The Fourier-transform takes the equation (\ref{eq:FRVghost}) of the ghost modes into
\begin{equation}
  \label{eq:ghost-four}
  \mathcal{D}_3 \chi = \Omega^2 \chi,
\end{equation}
while a gauge transformation takes the form
\begin{equation}
  \label{eq:gauge-four}
\delta A_i \to \delta A_i + \partial_i \chi,
\qquad \delta A_3 \to \delta A_3 - i k \chi, \qquad \delta A_0 \to \delta A_0 + i\Omega \chi.
\end{equation}
It is useful to introduce complex coordinates
\begin{equation}
  \label{eq:komplexkoord1}
  A_+ = \frac{e^{-i\vartheta}}{\sqrt{2}}(A_r - \frac{i}{r}A_\vartheta) \hspace{2.5cm} A_+ = \frac{e^{i\vartheta}}{\sqrt{2}}(A_r + \frac{i}{r}A_\vartheta).
\end{equation}
Fourier expansion in the angle variable  (in cylindrical coordinates $x^1=r, x^2=\vartheta$, see Eq.\ (\ref{eq:Fourtheta})),
omitting the sum over $\ell$ yields
\begin{equation}
  \label{eq:FRVpertans}
  \begin{aligned}
    \delta\phi_1 (k,\Omega)  &= s_{1,\ell} e^{i(n+\ell)\vartheta}\,, \\
    \delta\phi_2 (k,\Omega)  &= s_{2,\ell} e^{i(m+\ell)\vartheta}\,, \\
    \delta A_+   (k,\Omega)  &= i a_\ell e^{i(\ell-1)\vartheta}\,, \\
    \delta A_0   (k,\Omega)  &= a_{0,\ell} e^{i\ell\vartheta}\,,\\
    \delta A_3   (k,\Omega)  &= a_{3,\ell} e^{i\ell\vartheta}\,,\\
  \end{aligned}
  \quad\quad\quad
  \begin{aligned}
    \delta\phi_1^* (-k,-\Omega) &= s_{1,-\ell}^* e^{-i(n-l)\vartheta}\,, \\
    \delta\phi_2^* (-k,-\Omega) &= s_{2,-\ell}^* e^{-i(m-l)\vartheta}\,, \\
    \delta A_-   (-k,-\Omega)     &= -i a_{-\ell}^*  e^{i(\ell+1)\vartheta}\,, \\
    a_{0,-\ell}^* (-k,-\Omega)    &= a_{0,\ell}(k,\Omega)\,, \\
    a_{3,-\ell}^* (-k,-\Omega)    &= a_{3,\ell}(k,\Omega)\,. \\
  \end{aligned}
\end{equation}
Substituting this into the equations of motion, (\ref{eq:FRVlag2b}) assumes the form of an eigenvalue problem (\ref{eq:FRVell}) with
the operator
\begin{equation}
  M_\ell=
  \label{eq:FRVpertEig}
  \begin{pmatrix}
D_1  & U_1   & V    & V'    & A_1  & A_1'  & B_1  & 0  \\
U_1  & D_1^* & V'   & V     & A_1' & A_1   & B_1  & 0  \\
V    & V'    & D_2  & U_2   & A_2  & A_2'  & B_2  & 0  \\
V'   & V     & U_2  & D_2^* & A_2' & A_2   & B_2  & 0  \\
A_1  & A_1'  & A_2  & A_2'  & D_3  & 0     & 0    & 0  \\
A_1' & A_1   & A_2' & A_2   & 0    & D_3^* & 0    & 0  \\
B_1  & B_1   & B_2  & B_2   & 0    & 0     & D_4  & 0  \\
0    & 0     & 0   & 0     & 0    & 0     & 0    & D_4
  \end{pmatrix}
\end{equation}
with
\begin{equation}
  \begin{aligned}
    D_1    &= D_s + \frac{(n(1-a)+\ell)^2}{r^2} + k^2 + 2 k \omega a_3 + W_1 \\
    D_1^*  &= D_s + \frac{(n(1-a)-\ell)^2}{r^2} + k^2 - 2 k \omega a_3 + W_1 \\
    D_2    &= D_s + \frac{(m-na+\ell)^2}{r^2} + (k-\omega)^2 + 2 (k-\omega) \omega a_3 + W_2 \\
    D_2^*  &= D_s + \frac{(m-na-\ell)^2}{r^2} + (k+\omega)^2 - 2 (k+\omega) \omega a_3 + W_2 \\
    D_3    &= D_a +\frac{(\ell-1)^2}{r^2} \\
    D_3^*  &= D_a +\frac{(\ell+1)^2}{r^2} \\
    D_4    &= D_a +\frac{\ell^2}{r^2} \\
\end{aligned}\end{equation}
with
\[\begin{aligned}
D_s &= -\nabla_r^2+\omega^2 a_3^2+\beta(|f|^2-1)\,, \\
W_i &= (\beta+1)f_i^2\,, \\
D_a &= -\nabla_r^2+k^2 + 2|f|^2
\end{aligned}
\]
and
\[\begin{aligned}
  U_1   &= (\beta-1) f_1^2                                  \\
  V\,   &= (\beta+1) f_1 f_2                                \\
  A_1   &= -\sqrt{2}\left(f_1'-\frac{n f_1}{r}(1-a)\right)  \\
  A_1'  &= \sqrt{2}\left(f_1'+\frac{n f_1}{r}(1-a)\right)   \\
  B_1   &= 2\omega a_3 f_1
\end{aligned}\hspace{4cm}\begin{aligned}
  U_2   &=  (\beta-1) f_2^2 \\
  V'    &=  (\beta-1) f_1 f_2 \\
  A_2   &= -\sqrt{2}\left(f_2' -\frac{m-na}{r}f_2\right) \\
  A_2'  &= \sqrt{2}\left(f_2' +\frac{m-na}{r}f_2\right) \\
  B_2   &= 2 \omega (a_3-1) f_2 .
\end{aligned}
\]
The expansion of the gauge transformation generator function can be chosen as
\begin{equation}
  \label{eq:chi-four}
  \chi = \chi_\ell e^{i\ell\vartheta}.
\end{equation}
Using this expansion, the ghost mode equation (\ref{eq:ghost-four}) assumes the form
\begin{equation}
  \label{eq:ghost-ell}
  D_4 \chi_\ell = \Omega^2 \chi_\ell.
\end{equation}
Gauge transformations satisfying the above equation act on the fields as
\begin{equation}
  \label{eq:ell-gtrf}
  \begin{aligned}
    s_{a,\ell} & \to  s_{a,\ell} + i \chi_\ell f_a\,, \\
    a_\ell    & \to  a_\ell - \frac{i}{\sqrt{2}}\left( \chi_\ell' +\frac{\ell\chi_\ell}{r} \right)\,, \\
    a_{3,\ell} & \to a_{3,\ell} - i k \chi_\ell\,,
  \end{aligned}
  \hspace{3.5cm}
  \begin{aligned}
    s_{a,-\ell}^* & \to  s_{a,-\ell}^* - i \chi_\ell f_a\,, \\
    a_{-\ell}^*   & \to  a_{-\ell}^* + \frac{i}{\sqrt{2}}\left( \chi_\ell' -\frac{\ell\chi_\ell}{r} \right)\,, \\
    a_{0,\ell} & \to a_{3,\ell} + i \Omega \chi_\ell\,.
  \end{aligned}
\end{equation}

\subsection{The $\beta\to\infty$ limiting case}
In the $\beta\to\infty$ case the linearized form of the
constraint $\phi_a^*\phi_a=1$ is
\begin{equation}
  \label{eq:LinCons}
  \phi_a^*\delta\phi_a + \phi_a\delta\phi_a^* =0.
\end{equation}
This can be taken into account by the substitution
\begin{equation}
  \label{eq:BVPans}
  \begin{aligned}
    s_{1,\ell} &= t_\ell \cos\theta + t_{1,\ell}, \\
    s_{2,\ell} &= -t_\ell \sin\theta + t_{2,\ell}, \\
  \end{aligned}\qquad
  \begin{aligned}
    s_{1,-\ell}^* &= t_\ell \cos\theta - t_{1,\ell}, \\
    s_{2,-\ell}^* &= -t_\ell \sin\theta - t_{2,\ell}, \\
  \end{aligned}
\end{equation}
and the equations of $t_\ell$, $t_{1,\ell}$ and $t_{2,\ell}$ can be obtained with a somewhat lengthy but straightforward calculation. Here we
present only the $\ell=0$ case:
\begin{equation}
  \label{eq:EQt0}
  \frac{1}{r}(r t_0')' = \frac{A_{t0}}{r^2} t_0 +D_{t0} t_0 + E_{t0} t_1 + F_{t0} t_2-\frac{m-n}{\sqrt{2}r}\sin(2\theta)(a_0+a_0^*)
  +\omega\sin(2\theta)a_{3,0}
\end{equation}
with $A_{t0}=-(m-n)(m+n-2na)\cos(2\theta)$, $D_{t0}=k^2-\Omega^2-\omega^2(1-2a_3)\cos(2\theta)$, $E_{t0}=2k\omega a_3\cos\theta$ and
$F_{t0}=2k\omega(1-a_3)\sin\theta$. Similarly
\begin{equation}
  \label{eq:EQt1}
  \frac{1}{r}(r t_1')' = \frac{A_{t1}}{r^2} t_1 +C_{t1} t_1 + D_{t1} t_0 + E_{t1} t_2-\sqrt{2}\cos\theta\theta'(a_0-a_0^*)
\end{equation}
with $A_{t1}=-\frac{1}{2}\left[m^2-n^2-4(m-n)na\cos^2\theta+(m^2-n^2)\cos(2\theta)\right]$, $C_{t1}=1+k^2-\Omega^2-\frac{\omega^2}{2}
-\cos(2\theta)+2\omega^2a_3\cos^2\theta-\frac{\omega^2}{2}\cos(2\theta)-(\theta')^2$, $D_{t1}=2k\omega a_3\cos\theta$ and $E_{t1}=\sin(2\theta)$, and
\begin{equation}
  \label{eq:EQt2}
  \frac{1}{r}(r t_2')' = \frac{A_{t2}}{r^2} t_2 +C_{t2} t_2 + D_{t2} t_0 + E_{t2} t_1+\sqrt{2}\sin\theta\theta'(a_0-a_0^*)
\end{equation}
with $A_{t2}=-\frac{1}{2}\left[n^2-m^2+4(m-n)na\cos^2\theta+(m^2-n^2)\cos(2\theta)\right]$, $C_{t2}=1+k^2-\Omega^2+\frac{\omega^2}{2}
+\cos(2\theta)-2\omega^2a_3\sin^2\theta-\frac{\omega^2}{2}\cos(2\theta)-(\theta')^2$, $D_{t2}=-2k\omega (a_3-1)\sin\theta$ and $E_{t2}=\sin(2\theta)$.

\subsection{The expansion of the perturbation operator in the bifurcation parameter}\label{app:perteqnsSO}
Let us first expand the components of the perturbation matrix in Eq.\ (\ref{eq:FRVpertEig}) to second order in $\epsilon$.
Here we present only the terms used later on in this paper.
The differential operators expand as
\begin{equation}\label{eq:D2eps}
\begin{aligned}
D_2^{(0)} &= -\nabla_r^2 + \frac{(m-n a^{(0)}+\ell)^2}{r^2} + (k-\omega_{\rm b})^2 + \beta( (f_1^{(0)})^2 -1)\,, \\
D_2^{(1)} &= 0\,, \\
D_2^{(2)} &= 2\omega_{\rm b}\omega_2 -2k\omega_2 -\frac{2(m-na^{(0)}+\ell)na^{(2)}}{r^2} + 2\omega_{\rm b}(k-\omega_{\rm b})a_3^{(2)}
+ 2\beta f_1^{(0)} f_1^{(2)} +(2\beta+1) (f_2^{(1)})^2\,,
\end{aligned}
\end{equation}


\begin{equation}\label{eq:D2ceps}
\begin{aligned}
D_2^{*(0)} &= -\nabla_r^2 + \frac{(m-n a^{(0)}-\ell)^2}{r^2} + (k+\omega_{\rm b})^2 + \beta( (f_1^{(0)})^2 -1)\,, \\
D_2^{*(1)} &= 0\,, \\
D_2^{*(2)} &= 2\omega_{\rm b}\omega_2 +2k\omega_2 -\frac{2(m-na^{(0)}-\ell)na^{(2)}}{r^2} - 2\omega_{\rm b}(k+\omega_{\rm b})a_3^{(2)}
+ 2\beta f_1^{(0)} f_1^{(2)} +(2\beta+1) (f_2^{(1)})^2\,,
\end{aligned}
\end{equation}
Scalar-scalar interaction terms:
\begin{equation}
\label{eq:UV1eps}
\begin{aligned}
  V^{(0)}   &= 0\,,\\
  {V'}^{(0)}   &= 0\,,\\
\end{aligned}\qquad
\begin{aligned}
  V^{(1)}   &= (\beta+1)f_1^{(0)} f_2^{(1)}\,,\\
  {V'}^{(1)}   &= (\beta-1)f_1^{(0)} f_2^{(1)}\,,\\
\end{aligned}\qquad
\begin{aligned}
  V^{(2)} &= 0 \,,\\
  {V'}^{(2)} &= 0 \,.\\
\end{aligned}
\end{equation}
Scalar-vector interaction terms:
\begin{equation}
  \label{eq:A2eps}
\begin{aligned}
  A_2^{(1)} &= -\sqrt{2} \left({f_2^{(1)}}'-\frac{m-na^{(0)}}{r}f_2^{(1)}\right)\,,\\
  {A_2'}^{(1)} &= \sqrt{2} \left({f_2^{(1)}}'+\frac{m-na^{(0)}}{r}f_2^{(1)}\right)\,,\\
\end{aligned}
\end{equation}
while $A_2^{(0)}={A_2'}^{(0)}=A_2^{(2)}={A_2'}^{(2)}=B_2^{(0)}=B_2^{(2)}=0$, and
\begin{equation}
  \label{eq:B2eps}
  B_2^{(1)} = -2\omega_{\rm b} f_2^{(1)}\,.
\end{equation}
This $\epsilon$-expansion of the perturbation matrix elements can be used to obtain the unstable eigenvalue of twisted vortices
close to the bifurcation point via perturbation series.

\section{Perturbations of the ANO vortices}\label{sec:local}
The equations describing local vortices can be obtained from the twisted vortex equations (\ref{eq:FRVprofE}) by taking
$f_2=a_3=0$. With the same numerical methods as used in the case of the twisted vortices, local vortex background data can be obtained,
see Table \ref{tab:lokvBG}. Here, the $f^{(1)}$ and $a^{(2)}$ is defined by the behavior of the vortex profile functions at the origin,
$f\sim f^{(1)}r+O(r^2)$ and $a\sim a^{(2)}r^2 + O(r^4)$ (shooting parameters).

\begin{table}[!ht]
\begin{center}\begin{tabular}{|c|c|c|}
\hline
$\beta$ & $f^{(1)}$ & $a^{(2)}$ \\
\hline\hline
1.25  & 0.92418 &  0.53485 \\
2     & 1.09935 &  0.61657 \\
2.5   & 1.19677 &  0.65959 \\
\hline
\end{tabular}\end{center}
\caption{Local vortex background data}\label{tab:lokvBG}
\end{table}

\begin{figure}[!ht]
\noindent\hfil\includegraphics[scale=.3,angle=-90]{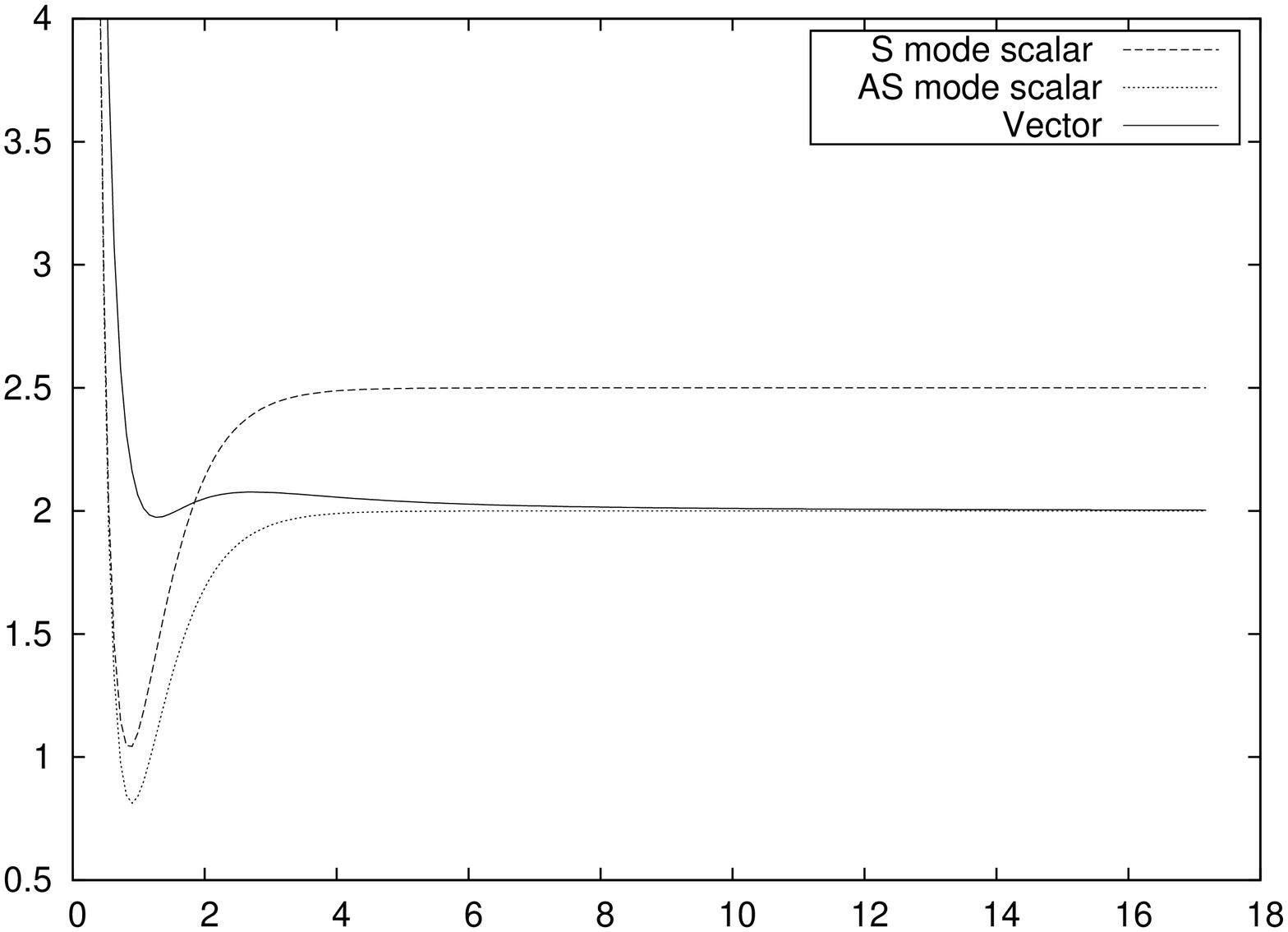} \includegraphics[scale=.3,angle=-90]{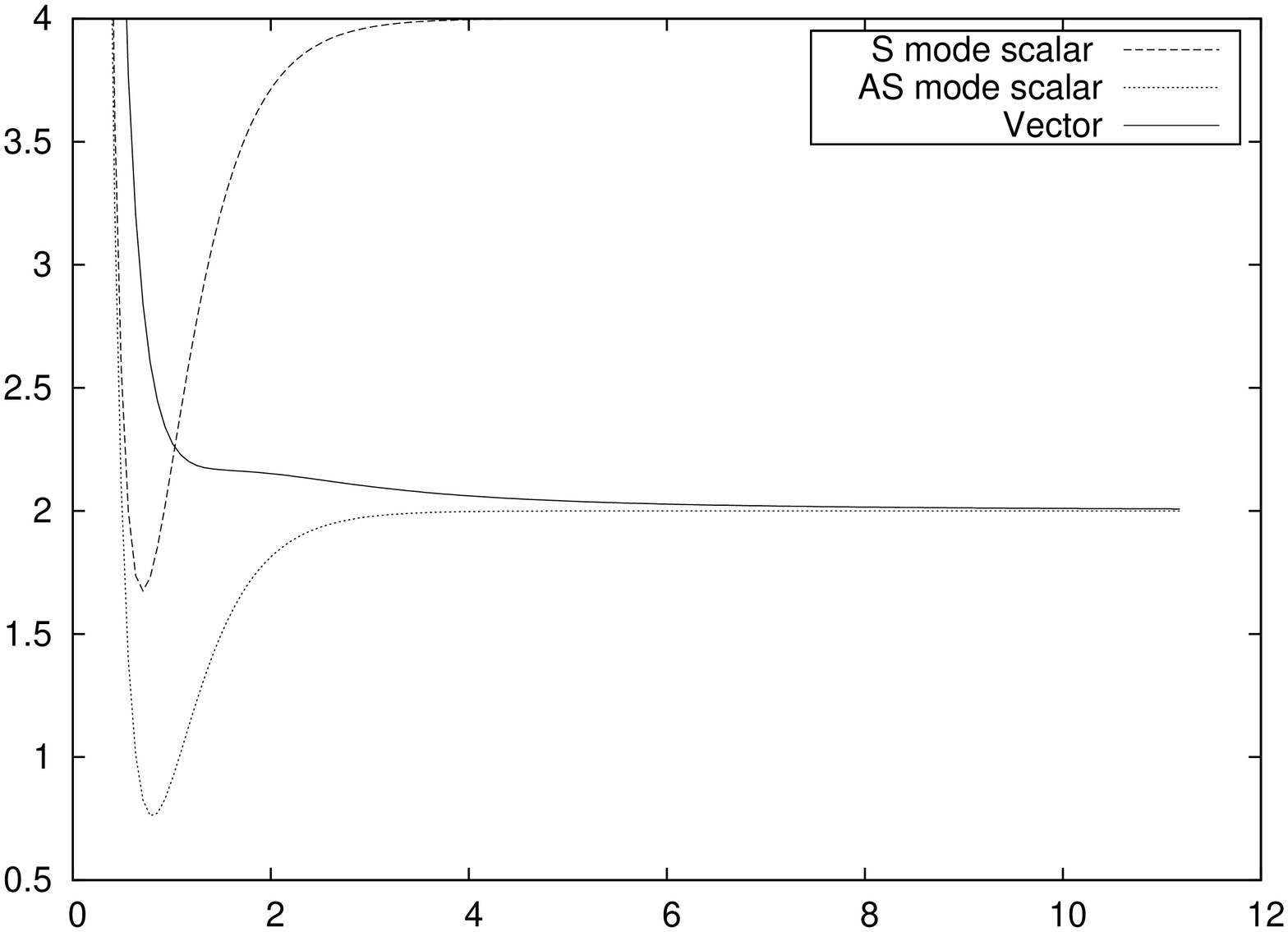}

\noindent\hfil\includegraphics[scale=.3,angle=-90]{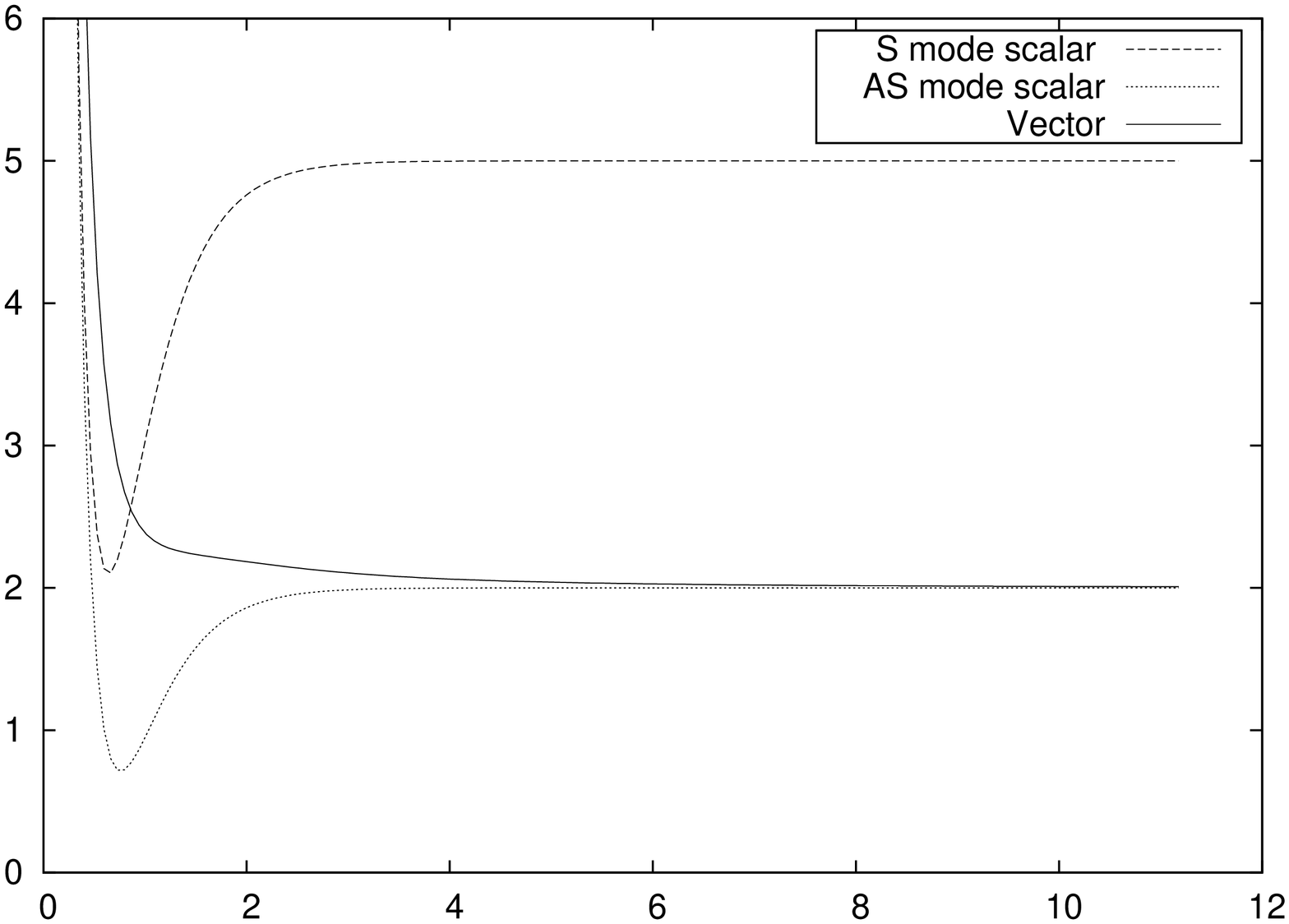}
\caption{Potentials in the ANO vortex perturbation equations for $\beta=1.25$ (left), $\beta=2$ (right) and $\beta=2.5$ (below)}
\label{fig:ANO-pot}
\end{figure}

\begin{table}[!ht]
\begin{center}\begin{tabular}{|c||c|c|c|}
\hline
         & Symm. mode & Anti. mode & $a_3$ mode \\
\hline
$\beta$  & $\lambda_S^2$ & $\lambda_A^2$ & $\lambda_{a3}^2$ \\
\hline\hline
1.25     & 1.82404                & 1.62442                & 1.62442 \\
2        &                        & 1.76100                & 1.76100 \\
2.5      &                        & 1.81813                & 1.81813 \\
\hline
\end{tabular}\end{center}
\caption{Parameters for the $\ell=0$ ANO vortex perturbation modes for $\beta=1.25$, $2$
and $2.5$}\label{tab:ANOpert}
\end{table}

The perturbations of local vortices can be described using equation (\ref{eq:FRVpertEig}) with $\omega=\omega_{\rm b}$ and $f_2=a_3=0$.
The $a_{3\ell}$ component decouples, and it's equation becomes that of the ghost modes, Eq.\ (\ref{eq:ghost-ell}). The $s_{2,\ell}$ and
$s_{2,-\ell}^*$ components also decouple; their equation (\ref{eq:bifur-f2}) has already been given in Section \ref{sec:vort}.
The instabilities arising from the extension of the model with a second Higgs field are described by these equations.
For further details see \cite{hin, FRV}.
In the $\ell=0$ case, there is a further decoupling in the $(s_{1,0},s_{1,0}^*,a_{0},a_{0}^*)$ sector: symmetric:
$\psi_S=(s_1,s_1,a_0,a_0)$ and antisymmetric $\psi_A=(s_1,-s_1,a_0,-a_0)$ modes are decoupled. For values of $\beta<1.5$ there is a bound mode
in each of these two sectors. For $\beta>1.5$ the symmetric mode becomes quasi-bound. The parameters for these modes (normed in a way that
$s_1'(0)=1$) are summarized in Table \ref{tab:ANOpert}. There, $\lambda_{A,S,a3}^2$ denotes the lowest eigenvalue (corresponding to $k=0$).
Figure \ref{fig:ANO-pot} shows the potential terms for $s_0$ and $a_0$ in the
symmetric and antisymmetric sectors, respectively, for $\beta=1.25$, $2$ and $2.5$. These explain it clearly why the mode $\psi_S$ becomes
quasi-bound for higher $\beta$ (i.e.\ for $\beta >1.5$, there is no bound mode in the symmetric sector, but for $2 < \Omega^2 < 2\beta$, the scalar
channel of the scattering problem is closed).

The gauge dependence of the eigenmodes and the relation of ghost modes to these is rather straightforward in the case of the ANO vortices.
In the analysis of the spectrum of these vortices, in Ref.\ \cite{Goodband} it has been
noted that if one considers the perturbation
problem of the vortices in 2+1D (i.e.\ $z$-independent perturbations), then the ghosts cancel the perturbations of $A_0$, and some modes of
the gauge fixed perturbation operators. If we allow $z$-dependent perturbations, instead of $\delta A_0$ a combination of $\delta A_0$ and
$\delta A_3$ is canceled. Ref.\ \cite{Goodband} also identifies the non-physical mode with the antisymmetric mode in the $\ell=0$ case,
based on the small difference in their numerical eigenvalues. Here, we would like to supplement this identification of the canceled mode
with some analytical calculation. Examining Eqns.\ (\ref{eq:ell-gtrf}), one can easily conclude that in the symmetric case, the
gauge generator drops out of $s_1 = (s_{1,0}+s_{1,0}^*)/2$ and $a_0 = (a_0+a_0^*)/2$. On the contrary, such a cancellation does not happen
in the antisymmetric case, thus the ghost eigenvalues shall appear in the spectrum of the antisymmetric mode. On a twisted vortex background,
the couplings are more complicated, and therefore such a straightforward analysis is not possible.


\begin{thebibliography}{999}
%
\newcommand{\NPB}{\sl Nucl.\ Phys.\ \bf B\,}
\newcommand{\PLB}{\sl Phys.\ Lett.\ \bf B\,}
\newcommand{\PL}{\sl Phys.\ Lett.\ \bf}
\newcommand{\PRD}{\sl Phys.\ Rev.\ \bf D\,}
\newcommand{\PRL}{\sl Phys.\ Rev.\ Lett.\ \bf}
\newcommand{\PRe}{\sl Phys.\ Rept.\ \bf}
\newcommand{\SJNP}{\sl Sov.\ J.\ Nucl.\ Phys.\ \bf}
\newcommand{\RPP}{\sl Rep.\ Prog.\ Phys.\ \bf}
\newcommand{\CMP}{\sl Commun.\ Math.\ Phys.\ \bf}
\newcommand{\ZPhys}{\sl Zeitschr.\ Phys.\ \bf}

\bibitem{cs-review}
A.~Vilenkin and E.~P.~S.~Shellard, Cosmic Strings and Other Topological Defects,
C.U.P. Cambridge, (1994);
M.~Hindmarsh and T.~W.~B.~Kibble, {\RPP 58} (1995) 477.
%
%
\bibitem{vac-ach} T.~Vachaspati and A.~Ach\'ucarro, {\PRD 44} (1991) 3067.
%
\bibitem{hin}  M.~Hindmarsh, {\PRL 68} (1992) 1263;
{\NPB 392} (1993) 461-492.
%
\bibitem{Vach91} T. Vachspati, {\PRL 68} (1991) 1977.
%
\bibitem{semilocal}
A.~Ach\'ucarro and T.~Vachaspati, {\PRe 327} (2000) 427.
%
\bibitem{bog}
E.~B.~Bogomol'nyi, {\SJNP 24} (1976) 449;
H.~J.~de Vega and F.~A.~Schaposnik, {\PRD 14} (1976) 1100.
%
\bibitem{Preskill}
J.~Preskill, {\PRD 46} (1992) 4218.
%
\bibitem{instab}
M.~James, L.~Perivolaropoulos and T.~Vachaspati, {\NPB 395} (1993) 534--546;
W.~B.~Perkins, {\PRD 47} 5224 (1993);
A.~Ach\'ucarro, R.~Gregory, J.A.~Harvey and K.~Kuijken, {\PRL 72} (1994) 3467;
M.~ Goodband, M.~Hindmarsh, {\PLB 363} (1995) 58.

%
\bibitem{FRV}
{Forg\'acs, P., Reuillon, S. and Volkov, M.S.}, {\NPB 751} (2006) 390--418.
%
\bibitem{akpv} A.~Ach\'ucarro, K.~Kuijken, L.~Perivolaropoulos and T.~Vachaspati, {\NPB 388} (1992) 435--456.
%
\bibitem{Goodband}
M.~Goodband and M.~Hindmarsh, {\PRD 52} (1995) 4621.
%
%
\bibitem{FM}  P.~Forg\'acs and N.S.~Manton {\CMP 72} (1980) 15.
%
%
\bibitem{Landau}
L.D.~Landau and E.M.~Lifshitz. A course of Theoretical Physics I. Mechanics, Pergamon Press, Oxford, 1976.
%
\bibitem{LandauQM}
L.D.~Landau and E.M.~Lifshitz. A course of Theoretical Physics III. Quantum Mechanics, Pergamon Press, Oxford, 1977.
%
\bibitem{numrec}
W.~H. Press, S.~A. Teukolsky, W.~T. Vetterling, and B.~P. Flannery. Numerical Recipes in C++, Cambridge University Press, 2002.

\bibitem{volkov}
J.~Garaud and M.S.~Volkov, {\NPB 799} (2008) 430--455.

\bibitem{sphaleron}
J.~Kunz, B.~Kleihaus and Y.~Brihaye, {\PRD 46 } (1992) 3587.

\bibitem{nambu}
Y.~Nambu, {\NPB 130} (1977) 505.


\bibitem{Nambu2}
Y.~Nambu, {\PRD 10} (1974) 4262.

\bibitem{preskill}
J.~Preskill, {\PRD 46} (1992), 4218.



\bibitem{vachEW}
T.~Vachaspati, {\PRL 68} (1992) 1977.

\bibitem{SalomaaVolovik}
M.M.~Salomaa and G.E.~Volovik, {\PL 51} (1983) 2040, {\PL 56} (1986), 363.


\bibitem{Witten}
 E.~Witten, {\NPB 249} (1985) 557--592.

\bibitem{GHew}
M.~Goodband and M.~Hindmarsh, {\PLB 363} (1995) 58--64.

\bibitem{cosmic}
M.~Hindmarsh and T.W.B.~Kibble, {\RPP 58} (1995) 477--462.


\bibitem{Leese}
R.~Leese, {\PRD 46} (1992) 4677.

\bibitem{Lemperiere}
Y.~Lemperiere, E.P.S.~Shellard, {\NPB 649} (2003) 511--525.

\bibitem{vorton}
R.A.~Battye, P.M.~Sutcliffe, {\NPB 814} (2009) 180--194.

\end{thebibliography}
\end{document}